\documentclass[reprint, amsmath, amssymb, aps, nofootinbib]{revtex4-2}

\usepackage{graphicx}
\usepackage{dcolumn}
\usepackage{bm}
\usepackage{braket}
\usepackage{amsmath}
\usepackage{here}
\usepackage{color}

\usepackage[small]{caption}
\usepackage[subrefformat=parens]{subcaption}
\usepackage{graphicx} 

\begin{document}

\title{Color-magnetic correlations in SU(2) and SU(3) lattice QCD}
\author{Atsuya Tokutake, Kei Tohme, Hideo Suganuma}

\affiliation{Division of Physics and Astronomy, Graduate School of Science, Kyoto University, \\
Kitashirakawaoiwake, Sakyo, Kyoto 606-8502, Japan}

\date{\today} 

\begin{abstract}
We study the two-point field-strength correlation 
$g^2 \langle G_{\mu\nu}^a(s)G^b_{\alpha\beta}(s') \rangle$
in the Landau gauge 
in SU(2) and SU(3) quenched lattice QCD, as well as the gluon propagator $g^2 \langle A_\mu^a (s)A_\nu^b(s') \rangle$.
The Landau-gauge gluon propagator $g^2 \langle A_\mu^a (s)A_\mu^a(s') \rangle$   
is well described by the Yukawa-type function 
$e^{-mr}/r$ with $r\equiv |s-s'|$ 
for $r=0.1-1.0~{\rm fm}$
in both SU(2) and SU(3) QCD.
Next, motivated by color-magnetic instabilities in the QCD vacuum, we investigate  
the perpendicular-type color-magnetic correlation, 
$C_{\perp}(r) \equiv g^2\langle H_z^a(s)H_z^a(s + r \hat \perp)) \rangle$ ($\hat \perp$: unit vector on the $xy$-plane), 
and the parallel-type one,  
$C_{\parallel}(r) \equiv g^2 \langle H_z^a(s)H_z^a(s + r \hat \parallel) \rangle$ 
($\hat \parallel$: unit vector on the $tz$-plane). 
These two quantities reproduce 
all the correlation of 
$g^2\langle G^a_{\mu\nu}(s)G^b_{\alpha\beta}(s')\rangle$, 
due to the Lorentz and global SU($N_c$) color symmetries in the Landau gauge.
Curiously, the perpendicular-type color-magnetic correlation 
$C_{\perp}(r)$ is found to be always negative for arbitrary $r$, 
except for the same-point correlation at $r$=0. 
In contrast, the parallel-type color-magnetic correlation $C_{\parallel}(r)$
is always positive. 
In the infrared region of $r \gtrsim 0.4~{\rm fm}$, 
$C_{\perp}(r)$ and $C_{\parallel}(r)$ 
strongly cancel each other, 
which leads to a significant  cancellation in the sum of the field-strength correlations as 
$\sum_{\mu, \nu} g^2\langle G^a_{\mu\nu}(s)G^a_{\mu\nu}(s')\rangle 
\propto C_{\perp}(|s-s'|)+ C_{\parallel}(|s-s'|) 
\simeq 0$.
Finally, we decompose the field-strength correlation 
into quadratic, cubic and quartic terms of the gluon field $A_\mu$ in the Landau gauge.
For the perpendicular-type color-magnetic correlation $C_{\perp}(r)$,  
the quadratic term is always negative, which is explained 
by the Yukawa-type gluon propagator. 
The quartic term gives a relatively small contribution.
In the infrared region, 
the cubic term is positive and tends to cancel with the quadratic term, 
resulting in a small value of $C_{\perp}(r)$.
\end{abstract}
\maketitle

\section{Introduction}

Quantum chromodynamics (QCD) is established as the fundamental theory of the strong interaction, 
and it is described as 
an SU($N_c$) non-abelian gauge theory with the color number $N_c=3$
\cite{Nambu66,GW73,P73}. 
Stemming from its non-abelian nature, QCD shows the asymptotic freedom, that is, 
its coupling decreases with the renormalization scale.
Therefore, perturbative QCD 
\cite{M09} is workable for the analysis of 
high-energy hadron reactions 
in the framework of the parton model \cite{F69,BP69}. 

At low energies, however, 
the QCD coupling becomes strong, 
and perturbative QCD is no more applicable. 
Reflecting the strong-coupling nature in the infrared region, 
QCD exhibits nonperturbative phenomena such as 
color confinement \cite{G03}
and dynamical chiral symmetry breaking \cite{NJ61}, 
and the QCD vacuum structure itself becomes highly nontrivial \cite{Handbook_2023}. 

For instance, due to the asymptotic freedom, QCD has a color-magnetic instability, which induces spontaneous emergence of color-magnetic fields \cite{S77,MS78,MSS81,S84,S25,NO79,AO80}. In other words, the system with
zero color-magnetic field is energetically unstable. The non-zero color-magnetic QCD vacuum is
called the Savvidy vacuum and/or the Copenhagen vacuum. Actually, the gluon condensate takes a large positive value, i.e., 
$\langle G_{\mu\nu}^a G^{\mu\nu}_a\rangle = 
2(\langle H_a^2 \rangle-
\langle E_a^2 \rangle) > 0$, 
in QCD in the Minkowski space, which means significant excess of color-magnetic fields rather than color-electric fields.

Historically, Savvidy first showed 
the magnetic instability in the Yang-Mills theory in 1977, using the one-loop effective potential with constant $H$, that is, 
the system with a constant color-magnetic field $H$ 
has lower energy than that of the perturbative vacuum with $H=0$ \cite{S77}. 
In fact, the energy density $\varepsilon(H)$ of 
a constant color-magnetic field 
is given by 
\begin{eqnarray}
\varepsilon(H)-\varepsilon(0)&=&\frac12 H^2 \cr
&+&\frac{11(gH)^2}{48\pi^2}
\left({\rm ln}\frac{gH}{\mu^2}-\frac12\right)-i\frac{(gH)^2}{8\pi}
\end{eqnarray}
with the gauge coupling $g$ at the renormalization point $\mu$
in the SU(2) Yang-Mills theory at the one-loop level.

Here, the logarithmic term includes the 
$\beta$-function coefficient with a minus sign, 
and this term causes instability of 
the $H=0$ system 
in the case of asymptotic freedom.
The energy minimum is achieved at non-zero color-magnetic field as
\begin{eqnarray}
gH=\mu^2\exp\Big(-\frac{24\pi^2}{11g^2}\Big).
\end{eqnarray}
This color-magnetic instability is logarithmic-type 
and cannot be removed by higher-order corrections.

The solution of the one-loop effective action with $H(x)$ is known to be a multi-vortex system, 
which was found by the Copenhagen group around 1980 \cite{AO80}.
The multi-vortex system is inhomogeneous and 
clearly breaks translational and rotational symmetry, 
while the physical vacuum is Poincare invariant. 

Then, at an infrared scale, such inhomogeneous systems must appear as randomly-directed domains to recover the rotational symmetry in the physical QCD vacuum \cite{NO79,MSS81,S84}, 
like magnetic materials above the Curie temperature. 
In this way, the multi-vortex system is conjectured to form  
a fluctuating stochastic domain structure at a large scale.
Considering the fluctuating color fields in the QCD vacuum \cite{MSS81,S84}, 
Kirzhnits et al. showed that a possible randomization of QCD can produce an effect similar to localization, resulting in a linear inter-quark potential \cite{AKL82}. 
Dosch and Simonov proposed the ``stochastic vacuum model" for gauge-invariant field-strength correlators and demonstrated that its infrared exponential damping yields an asymptotic linear potential \cite{D87,S88,DS88}. 

In spite of the importance of the color-magnetic instability, which is directly related to the QCD-vacuum structure, such analytical research has not been well studied beyond the one-loop effective action. 
One obvious reason is due to theoretical difficulties to analyze 
nonperturbative properties in infrared QCD.

Nowadays, lattice QCD \cite{C80,R12,MM94} is a standard theoretical method 
for the analysis of nonperturbative QCD, and it is numerically performed with 
the Monte Carlo method based on the Euclidean path integral of QCD.
In lattice QCD, as an importance sampling, 
lattice gauge configurations are numerically generated, and their set can be regarded as 
a typical ensemble of the QCD vacuum. 
Therefore, one obtains much important information of the nonperturbative-QCD vacuum 
by analyzing the lattice QCD configurations.

As a pioneering work, based on the stochastic vacuum model,  
Di~Giacomo~et~al. \cite{GP92,EGM97,GMP97,GDSS02,EGM03} and Bali, Brambilla and Vairo \cite{BBV98} found that the gauge-invariant field-strength correlator exhibits infrared exponential damping in lattice QCD.

Motivated by these studies, we study the field-strength correlation and its overall behavior in lattice QCD \cite{STT25}. 
In this paper, using lattice QCD, we mainly investigate the color-magnetic correlation in the Landau gauge, which has many advantages in terms of symmetries and minimal gauge-field fluctuations. 

The organization of this paper is as follows.
In Sec.~II, we present 
the formalism of color-magnetic correlations
in QCD in the Landau gauge.
Section~III summarizes the setup of lattice QCD calculations in the case of SU(2) and SU(3) color, respectively.
In Sec.~IV, we show the lattice QCD result of the Landau-gauge gluon propagator.
Section~V shows the numerical results of the color-magnetic correlations in SU(2) and SU(3) lattice QCD.
In Sec.~VI, we analyze the color-magnetic correlations 
in terms of the order of the gluon field. 
Section~VII is devoted for the summary and concluding remarks.

\section{Color-magnetic correlations in QCD}

This section presents the formalism of
the color-magnetic correlation 
$g^2\langle H^a_z(s)H^a_z(s')\rangle$,  
which is generalized to the field-strength correlation 
$g^2\langle G^a_{\mu\nu}(s)G^b_{\alpha\beta}(s') \rangle$, in the Landau gauge in SU($N_c$) QCD.

\subsection{Color-magnetic correlations in the Landau gauge}

SU($N_c$) QCD is described by 
the quark field
$q(s)$ and the gluon field $A_\mu(s)= A^a_\mu(s) T^a\in {\rm su}(N_c)$ ($T^a$: SU($N_c$) generators), 
and its Lagrangian is expressed as 
\cite{M09,R12}
\begin{eqnarray}
{\cal L}=-\frac12 {\rm Tr}~ G_{\mu\nu}G^{\mu\nu}+\bar q(i\not\!\! D-m)q.
\end{eqnarray}
The quark field $q_{c,f}^{\alpha}(s)$ has a color index $c=1,2,\cdots, N_c$ in the fundamental representation, 
as well as a Dirac spinor index 
$\alpha$ and a flavor index $f$.
We have used the covariant derivative $D_\mu \equiv \partial_\mu +igA_\mu$
and 
the field strength
\begin{eqnarray}
G_{\mu\nu}&\equiv&G_{\mu\nu}^aT^a\equiv
\frac1{ig}[D_\mu, D_\nu] \cr
&=&\partial_\mu A_\nu -
\partial_\nu A_\mu +
ig[A_\mu, A_\nu]
\in {\rm su}(N_c),
\end{eqnarray}
with the gauge coupling $g$.
The color-magnetic field $H_i(s)$ $(i=1,2,3)$ is given by 
the spatial field strength,
\begin{eqnarray}
    H_i(s) =H_i^a(s)T^a= \frac12 \epsilon_{ijk}G_{jk}(s) \in {\rm su}(N_c).
\end{eqnarray}

In this paper, we investigate 
the field-strength correlation 
$g^2\langle G^a_{\mu\nu}(s)G^b_{\alpha\beta}(s') \rangle$, 
which is generally gauge-variant in QCD, and its analysis requires some gauge fixing. (As an alternative, one could consider a gauge-invariant quantity similar to that by linking the two field strengths with suitable parallel transporters \cite{MM94}. However, this method gives uncertainty regarding how to make the connection.)

To investigate the correlation,  
we mainly use the Landau gauge 
because it is widely used due to its many useful symmetries, such as Lorentz, translational 
and global color symmetries. 
In Euclidean QCD, we define the Landau gauge 
so as to minimize the global quantity,
\begin{eqnarray}
R[A_\mu^a] \equiv 
\int d^4s~
\{A^a_\mu(s)A^a_\mu(s)\},
\end{eqnarray}
via a gauge transformation
\cite{MO87,ISI09,OS12,CMPS18},
which leads to the local gauge fixing condition
$\partial_\mu A_\mu=0$.
(Note that such a global definition cannot be applied in the Minkowski metric.)
In this global definition, the Landau gauge has a clear physical interpretation that it strongly suppresses in total artificial gauge-field fluctuations associated with the gauge degrees of freedom \cite{ISI09}.

Considering the nontrivial color-magnetic picture for the QCD vacuum, 
we investigate the following two types of color-magnetic correlation in the Landau gauge in lattice QCD \cite{STT25}:
\begin{enumerate}
\item 
Perpendicular-type color-magnetic correlation 
\begin{eqnarray}
C_{\perp}(r) \equiv g^2 \langle H^a_z(s)H^a_z(s+r \hat \perp)\rangle    
\end{eqnarray}
with $\hat \perp$ an arbitrary unit vector on the $xy$-plane, 
\item 
Parallel-type color-magnetic correlation 
\begin{eqnarray}
C_\parallel(r) \equiv g^2 \langle H^a_z(s)H^a_z(s+r \hat \parallel)\rangle 
\end{eqnarray}
with $\hat \parallel$ an arbitrary unit vector on the $tz$-plane.
\end{enumerate}

In Euclidean Landau-gauge QCD, 
we note the relation 
\begin{eqnarray}
\langle H^a_z(s)H^a_z(s+r\hat z)\rangle &=&\langle H^a_z(s)H^a_z(s+r\hat t)\rangle, \cr 
& i.e.&, \cr
\langle G_{xy}^a(s)G_{xy}^a(s+r\hat z)\rangle &=&\langle G_{xy}^a(s)G_{xy}^a (s+r\hat t)\rangle, 
\end{eqnarray}
due to the four-dimensional rotational invariance.
Furthermore, 
as shown in the next subsection, 
any field-strength correlation 
$g^2\langle G^a_{\mu\nu}(s)G^b_{\alpha\beta}(s') \rangle$ can be expressed with the two correlations 
$C_{\perp}(r)$ and $C_{\parallel}(r)$ 
in the Landau gauge.

\begin{figure}[htbp]
    \centering
    \includegraphics[width=5.6cm]{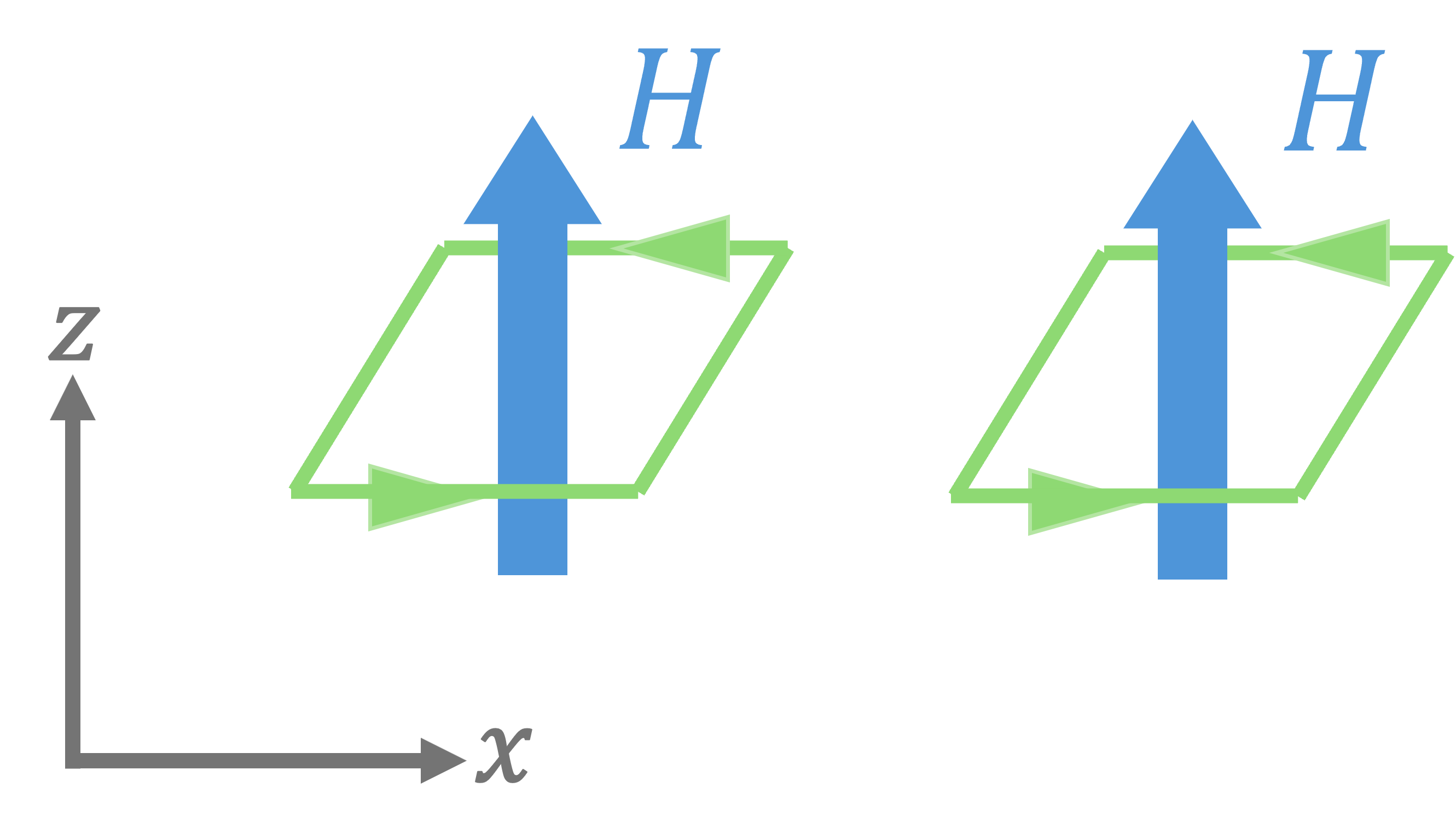}
    \includegraphics[width=4.2cm]{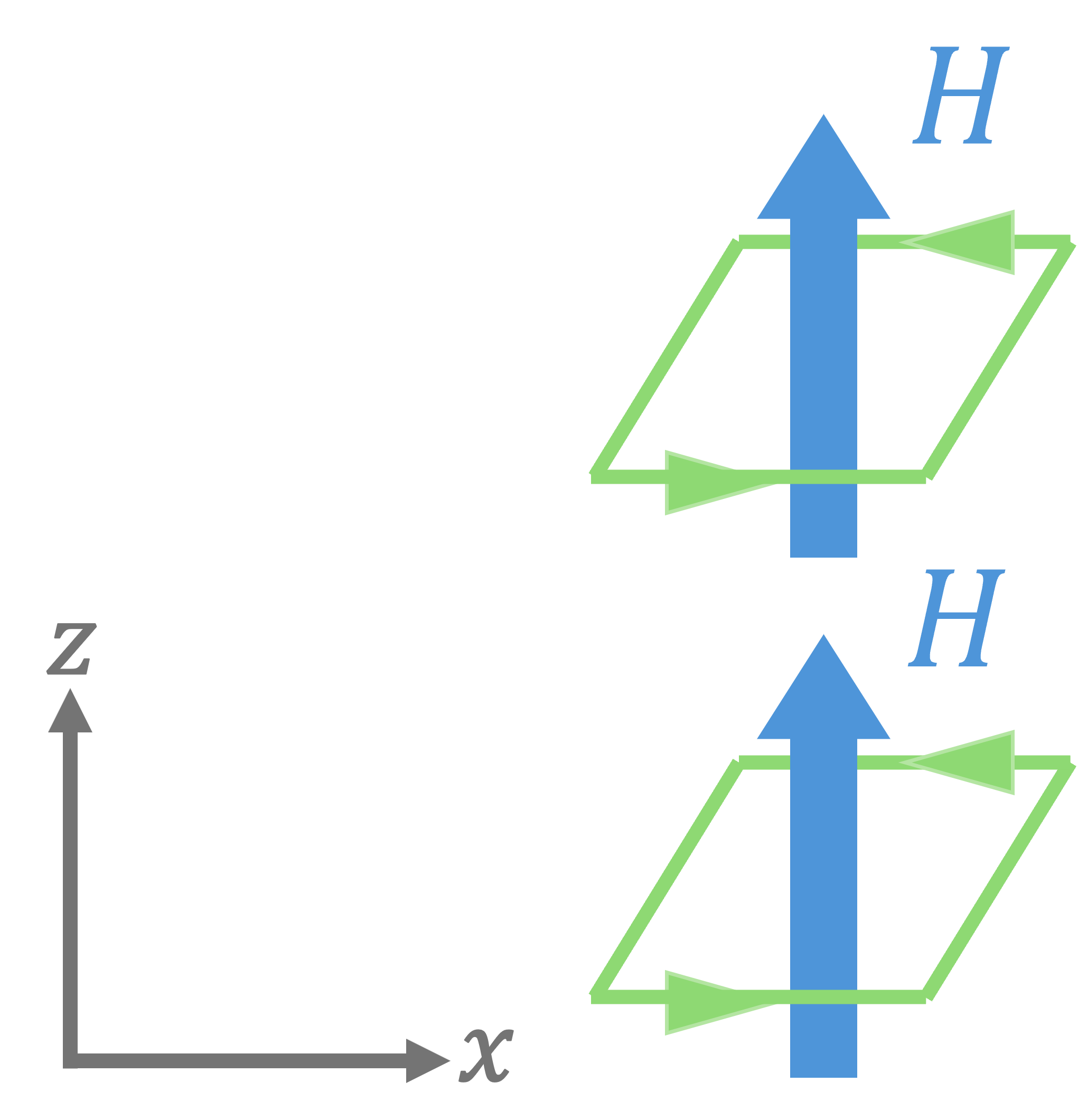}
    \caption{
Plaquette correlations corresponding to 
the spatial color-magnetic correlations at the same Euclidean time, 
$C_{\perp}(r)$ 
and $C_{\parallel}(r)$,
in lattice QCD. 
The upper plaquette correlation 
corresponds to 
$C_{\perp}(r) =
g^2\langle H_z^a(s) H_z^a(s+r\hat x) \rangle$, 
and the lower one $C_{\parallel}(r) =
g^2\langle H_z^a(s) H_z^a(s+r\hat z) \rangle$. 
}
    \label{fig:plaq-corr}
\end{figure}

In the lattice QCD formalism, 
$C_\perp(r)$ and $C_\parallel(r)$ 
correspond to 
the spatial correlation of 
two plaquettes, as illustrated in 
Fig.~\ref{fig:plaq-corr}.

\subsection{General field-strength correlation}

In Euclidean QCD in the Landau gauge, 
all two-point field-strength correlations  
$g^2\langle G^a_{\mu\nu}(s)G^b_{\alpha\beta}(s') \rangle$ can be expressed with the two correlations 
$C_{\perp}(r)$ and $C_{\parallel}(r)$,
due to the Lorentz and global SU($N_c$) color symmetries. 

Without loss of generality, we set 
\begin{eqnarray}
s^R\equiv s-s'=r\hat t= (0,0,0,r) ~~\hbox{(relatively temporal)}~~~~~
\end{eqnarray}
by a suitable 
four-dimensional rotation.
Then, because of color and reflection symmetries,
the field-strength correlation is written by  
\begin{eqnarray}
g^2\langle G^a_{\mu\nu}(s)G^b_{\alpha\beta}(s') \rangle    
=\frac{\delta^{ab}}{N_c^2-1}
(\delta_{\mu\alpha}\delta_{\nu\beta}-\delta_{\mu\beta}\delta_{\nu\alpha})K_{\mu\nu}(r) \cr \hbox{(no sum).~~~~~}
\label{eq:FScorr}
\end{eqnarray}
This quantity is antisymmetric 
with respect to the exchange of $\alpha$ and $\beta$.
Because it is also antisymmetric 
under the exchange of $\mu$ and $\nu$, 
$K_{\mu\nu}(r)$ has to be symmetric. 
Since the tensor factor in Eq.~(\ref{eq:FScorr}) 
becomes zero for $\mu=\nu$,  
the diagonal component of 
$K_{\mu\nu}(r)$ can be any value 
and is set to zero here. 
Then, $K_{\mu\nu}$ is expressed as
\begin{eqnarray}
K_{\mu\nu}(r)=K_{\nu\mu}(r), 
\quad K_{\mu\mu}(r)=0~~\hbox{(no sum)},
\end{eqnarray}
and one obtains from Eq.(\ref{eq:FScorr})
\begin{eqnarray}
K_{\mu\nu}(r)
=g^2\langle G_{\mu\nu}^a(s)G_{\mu\nu}^a(s')\rangle
~~\hbox{(no sum on $\mu$, $\nu$)}.
\end{eqnarray}

Due to the rotational symmetry, $K_{\mu\nu}(r)$ has 
only two independent components, 
\begin{eqnarray}
&&K_{tx}(r)=K_{ty}(r)=K_{tz}(r)=C_{\perp}(r)
\end{eqnarray}
and 
\begin{eqnarray}
&&K_{xy}(r)=K_{yz}(r)=K_{zx}(r)=C_{\parallel}(r).
\end{eqnarray}
Here, we have used the following relations 
based on four-dimensional rotation
symmetry in the Euclidean space-time: 
\begin{eqnarray}
K_{ty}(r)&=&g^2\langle 
G_{ty}^a(r\hat t)G_{ty}^a(0)\rangle
=g^2\langle 
G_{xy}^a(r\hat x)G_{xy}^a(0)\rangle \cr
&=&g^2\langle 
H_{z}^a(r\hat x)H_{z}^a(0)\rangle
=C_{\perp}(r)
\end{eqnarray}
by exchanging $t$ and $x$, 
and 
\begin{eqnarray}
K_{xy}(r)
&=&g^2\langle 
G_{xy}^a(r\hat t)G_{xy}^a(0)\rangle
=g^2\langle 
G_{xy}^a(r\hat z)G_{xy}^a(0)\rangle
\cr
&=&g^2\langle 
H_{z}^a(r\hat z)H_{z}^a(0)\rangle
=C_{\parallel}(r)
\end{eqnarray}
by exchanging $t$ and $z$.

In this way, we obtain for $s^R\equiv s-s'=r\hat t=(0,0,0,r)$ 
(relatively temporal)
\begin{widetext}
\begin{eqnarray}
g^2\langle G^a_{\mu\nu}(s)G^b_{\alpha\beta}(s') \rangle
&=&\frac{\delta^{ab}}{N_c^2-1}
\Big[\{\delta_{\mu 4}(\delta_{\alpha 4}\delta_{\nu\beta}-\delta_{\beta 4}\delta_{\nu \alpha})+[(\mu, \alpha) \leftrightarrow (\nu, \beta
)]\}C_{\perp}(r)
+\epsilon_{\mu\nu k 4}
\epsilon_{\alpha\beta k 4}C_{\parallel}(r)\Big]\cr
&=&\frac{\delta^{ab}}{N_c^2-1}
\left[\{\delta_{\mu 4}(\delta_{\alpha 4}\delta_{\nu\beta}-\delta_{\beta 4}\delta_{\nu \alpha})+
\delta_{\nu 4}(\delta_{\beta 4}\delta_{\mu\alpha}-\delta_{\alpha 4}\delta_{\mu \beta})
\}C_{\perp}(r)
+(\hat \delta_{\mu\alpha}\hat \delta_{\nu\beta}-\hat \delta_{\mu\beta}\hat\delta_{\nu\alpha})
C_{\parallel}(r)\right],~~~~~~
\label{eq:FSformula}
\end{eqnarray}
\end{widetext}
with $r=|s-s'|$ and $\hat\delta_{\mu\nu}
\equiv \delta_{\mu\nu}-\hat s^R_\mu \hat s^R_\nu
=\delta_{\mu\nu}-\hat t_\mu \hat t_\nu$. 
For the derivation, we have used its antisymmetric property under the exchange of $\mu \leftrightarrow \nu$
and $\alpha \leftrightarrow \beta$, respectively.
By applying a four-dimensional rotation to  
Eq.~(\ref{eq:FSformula}), 
all two-point field-strength correlations  
$g^2\langle G^a_{\mu\nu}(s)G^b_{\alpha\beta}(s') \rangle$ 
are expressed with $C_\perp(r)$ and 
$C_\parallel(r)$ in the Landau gauge.
In particular, one finds 
\begin{eqnarray}
\sum_{\mu, \nu}g^2 \langle G_{\mu\nu}^a(s) G_{\mu\nu}^a(s')\rangle
=6~[C_{\perp}(r)+C_{\parallel}(r)],
\end{eqnarray}
which is a single-valued function of 
$r=|s-s'|$ in the Landau gauge.

\section{gluon and field strength in SU(2) and SU(3) lattice QCD}

For the nonperturbative analysis 
of the color-magnetic correlation,
we use SU(2) and SU(3) lattice QCD Monte Carlo calculations 
with the standard plaquette action 
at the quenched level.
In lattice QCD, field variables are 
formulated on a lattice with the spacing $a$, 
and their interactions are controlled 
by the gauge coupling $g$ \cite{R12,MM94}.


In lattice QCD, 
the gluon field $A_\mu(s) \in {\rm su}(N_c)$ 
and 
the field strength $G_{\mu\nu}(s) \in {\rm su}(N_c)$ are exponentiated to form 
the link variable $U_\mu(s)$ and 
the plaquette $\square_{\mu\nu}(s)$, respectively: 
\begin{eqnarray}
U_\mu(s)&=&e^{iagA_\mu(s)}\in {\rm SU}(N_c), \cr
\square_{\mu\nu}(s)&=&e^{ia^2gG_{\mu\nu}(s)}\in {\rm SU}(N_c).
\end{eqnarray}

For the lattice data analysis, 
we use the translational and 
rotational invariance 
in Euclidean Landau-gauge QCD.
For the spatial correlation, 
we take both on-axis and 
off-axis lattice data.
On the statistical error of the
lattice data, 
the jackknife error estimate is adopted.

\subsection{SU(3) lattice QCD setup}

For the SU(3) lattice QCD calculations,  
we adopt the 
four lattices as shown in Table~\ref{tab:SU3setup}.

\begin{center}
\begin{table}[ht]
    \centering
    \begin{tabular}{ccccc} 
    \hline
    \hline
        ~$\beta$~ & ~ lattice size~ & ~ spacing $a$ ~ & ~ phys. vol. $(La)^4$ & ~ config.\# \\ 
    \hline
        5.7 & $16^4$ & ~0.186 fm & ~$(3.0~{\rm fm})^4$ &  200\\
        5.8 & $16^4$  & ~0.152 fm & ~$(2.4~{\rm fm})^4$ &  200\\
        6.0 & $24^4$  & ~0.104 fm & ~$(2.5~{\rm fm})^4$ &  800\\
        6.2 & $48^4$  & 0.0726 fm & $(3.48~{\rm fm})^4$ &  ~50\\
    \hline
    \hline
    \end{tabular}
    \caption{Lattice parameter $\beta \equiv 2N_c/g^2$ and the lattice size for the SU(3) lattice QCD calculations at the quenched level.
    The corresponding lattice spacing $a$ 
    and the physical volume $(La)^4$ are also listed.}
    \label{tab:SU3setup}
\end{table}
\end{center}

The lattice spacing $a$ 
is determined so as to reproduce the string tension 
$\sigma=0.89~{\rm GeV/fm}$ at $\beta$=5.7, 5.8 and 6.0 \cite{ISI09,TSNM02}.
For the lattice spacing at $\beta=6.2$, 
we refer Ref.~\cite{OS12}.
The gauge configurations are picked up with the interval of 1,000 sweeps, 
after the thermalization of 20,000 sweeps. 
In this study, we use 
200 gauge configurations  
at $\beta$=5.7 and 5.8, and 
800 gauge configurations at $\beta=6.0$.
For the gluon propagator, we also use 
50 gauge configurations at $\beta=6.2$. 
For the Landau gauge fixing, we use the ordinary iterative maximization algorithm \cite{MO87} with an over-relaxation parameter of 1.6.

In the Landau gauge, 
since the gluon field is globally minimized, 
we define SU(3) gluon fields 
with the link-variable as
\begin{eqnarray}
{\cal A}_\mu(s) \equiv \frac1{2iag}[U_\mu(s)-U^\dagger_\mu(s)]\Big|_{\rm traceless} \in {\rm su}(3)
\label{eq:SU3-gluon}
\end{eqnarray}
in the fundamental representation. 
This definition is often used 
in the Landau gauge in lattice QCD.
In the following,
we simply denote ${\cal A}_\mu(s)$ by $A_\mu(s)$ 
and mainly use it for the SU(3) lattice calculation.

In SU(3) lattice QCD, we define the field strength 
\begin{eqnarray}
G_{\mu\nu}(s) &\equiv& \partial_\mu A_\nu(s)
-\partial_\nu A_\mu(s)+ig[A_\mu(s), A_\nu(s)]
\cr
&\in& {\rm su}(3)
\end{eqnarray}
using the Landau-gauge 
gluon field in Eq.(\ref{eq:SU3-gluon})  
and 
the forward derivative, satisfying 
\begin{eqnarray}
\partial_\mu f(s)= \frac{1}{a} [f(s+a\hat \mu)-f(s)]. 
\end{eqnarray}
We mainly use the field strength $G_{\mu\nu}(s)$ for lattice QCD calculations of 
field-strength correlations.

We also introduce
SU(3) ``plaquette field strength'' 
\begin{eqnarray}
{\cal G}_{\mu\nu}(s) \equiv \frac1{2ia^2g}[\square_\mu(s)-\square^\dagger_\mu(s)]\Big|_{\rm traceless}\in {\rm su}(3)
\label{eq:P-elemag}
\end{eqnarray}
defined with the plaquette variable 
in the Landau gauge. 
${\cal G}_{\mu\nu}(s)$ coincides with 
$G_{\mu\nu}(s)$ 
in the continuum limit.

\subsection{SU(2) lattice QCD setup}


For the SU(2) lattice QCD calculations, 
we adopt the three lattices 
as shown in Table~\ref{tab:SU2setup}.
\begin{center}
\begin{table}[ht]
    \centering
    \begin{tabular}{ccccc} 
    \hline
    \hline
        ~$\beta$~ & ~ lattice size~ & ~ spacing $a$ ~ & ~ phys. vol. $(La)^4$ & ~ config.\# \\ 
    \hline
        2.3 & $16^4$ & ~0.18 fm & $(2.9~{\rm fm})^4$ & 400\\
        2.4 & $24^4$  & 0.127 fm & $(3.0~{\rm fm})^4$ & 400\\
        2.5 & $32^4$  & ~0.09 fm & $(2.9~{\rm fm})^4$ & 400\\
    \hline
    \hline
    \end{tabular}
    \caption{Lattice parameter $\beta \equiv 2N_c/g^2$ and the lattice size for the SU(2) lattice QCD calculations at the quenched level.
    The corresponding lattice spacing $a$ 
    and the physical volume $(La)^4$ are also listed.}
    \label{tab:SU2setup}
\end{table}
\end{center}

The lattice spacing $a$ is determined to reproduce the string tension 
$\sigma=0.89~{\rm GeV/fm}$ \cite{AS99}.
The gauge configurations are picked up with the interval of 200 sweeps, 
after the thermalization of 2,000 sweeps.
We use 400 gauge configurations at each $\beta$. 
%
%
The Landau gauge fixing is achieved 
by the ordinary iterative maximization algorithm \cite{MO87} with 
an over-relaxation parameter of 1.7.

In SU(2) lattice QCD, 
the gluon field 
$A_\mu(s) \in {\rm su}(2)$ is 
directly obtained from the 
link-variable $U_\mu(s)$ in the fundamantal representation as
\begin{eqnarray}
U_\mu(s)&=&e^{iagA_\mu(s)} 
=e^{iagA_\mu^a(s) \tau^a/2} \cr
&=&e^{i\tau^a\theta^a} 
=\cos\theta+i\tau^a\hat \theta^a \sin\theta 
\in {\rm SU}(2),
\label{eq:SU2-gluon}
\end{eqnarray}
with $\theta^a\equiv agA_\mu^a(s)/2$, 
%
$\theta \equiv (\theta^a\theta^a)^{1/2}$ and 
$\hat \theta^a \equiv \theta^a/\theta$.
In SU(2) lattice QCD, 
we define the field strength 
\begin{eqnarray}
G_{\mu\nu}(s)&\equiv& \partial_\mu A_\nu(s)
-\partial_\nu A_\mu(s)+ig[A_\mu(s), A_\nu(s)]\cr
&\in& {\rm su(2)}
\end{eqnarray}
using the Landau-gauge gluon field $A_\mu(s)$ in Eq.(\ref{eq:SU2-gluon})  
and the forward derivative $\partial_\mu$.

\section{Landau-gauge gluon propagator}

To begin with, 
we study the gluon propagator, which is closely related to the color-magnetic correlation. (Indeed, in abelian gauge theories, the gauge-field propagator and the field strength are directly related.) 
The gluon propagator is one of 
the most fundamental quantities in QCD \cite{M99,AS01,ABP08}
and has been well studied, particularly in the Landau gauge in lattice QCD \cite{ISI09,MO87,OS12,CMPS18}.

In this section, we numerically 
investigate the gluon propagator, 
i.e., the gluon two-point function, 
\begin{eqnarray}
&&D_{\mu\nu}^{ab}(s, s') \equiv 
g^2\langle A_\mu^a(s)A_\nu^b(s')\rangle=\frac{\delta^{ab}}{N_c^2-1}D_{\mu\nu}(s-s'),~~~~~
\\
&&D_{\mu\nu}(s-s') \equiv
g^2\langle A_\mu^a(s)A_\nu^a(s')\rangle,
\label{eq:propagator}
\end{eqnarray}
in the Landau gauge 
in Euclidean SU($N_c$) lattice QCD. 
In the Landau gauge, all components of 
the gluon propagator can be 
expressed with the scalar combination of 
\begin{eqnarray}
D(r)\equiv g^2 \langle A_\mu^a(s)A_\mu^a(s')\rangle,    
\end{eqnarray}
which is a single-valued function of 
the four-dimensional Euclidean space-time distance 
$r\equiv |s-s'|$ \cite{ISI09}.
We briefly summarize the general properties of 
the Landau-gauge gluon propagator in Appendix A.

Figure~\ref{fig:prop} shows 
the gluon propagator $D(r)$ 
in the Landau gauge 
in SU(3) and SU(2) quenched lattice QCD.
\begin{figure}[htbp]
    \centering
    \includegraphics[width=9.5cm]{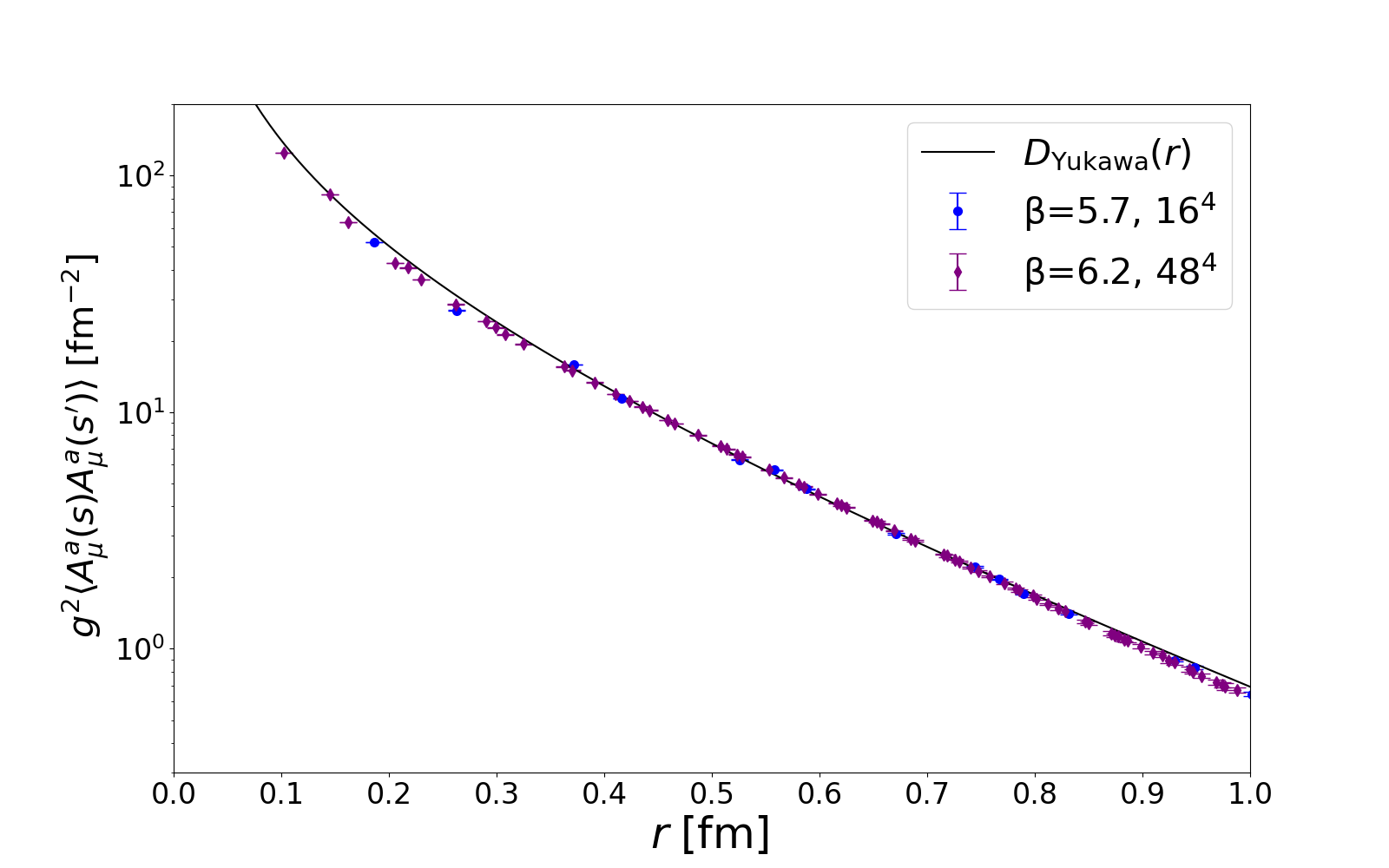}
    \includegraphics[width=9.5cm]{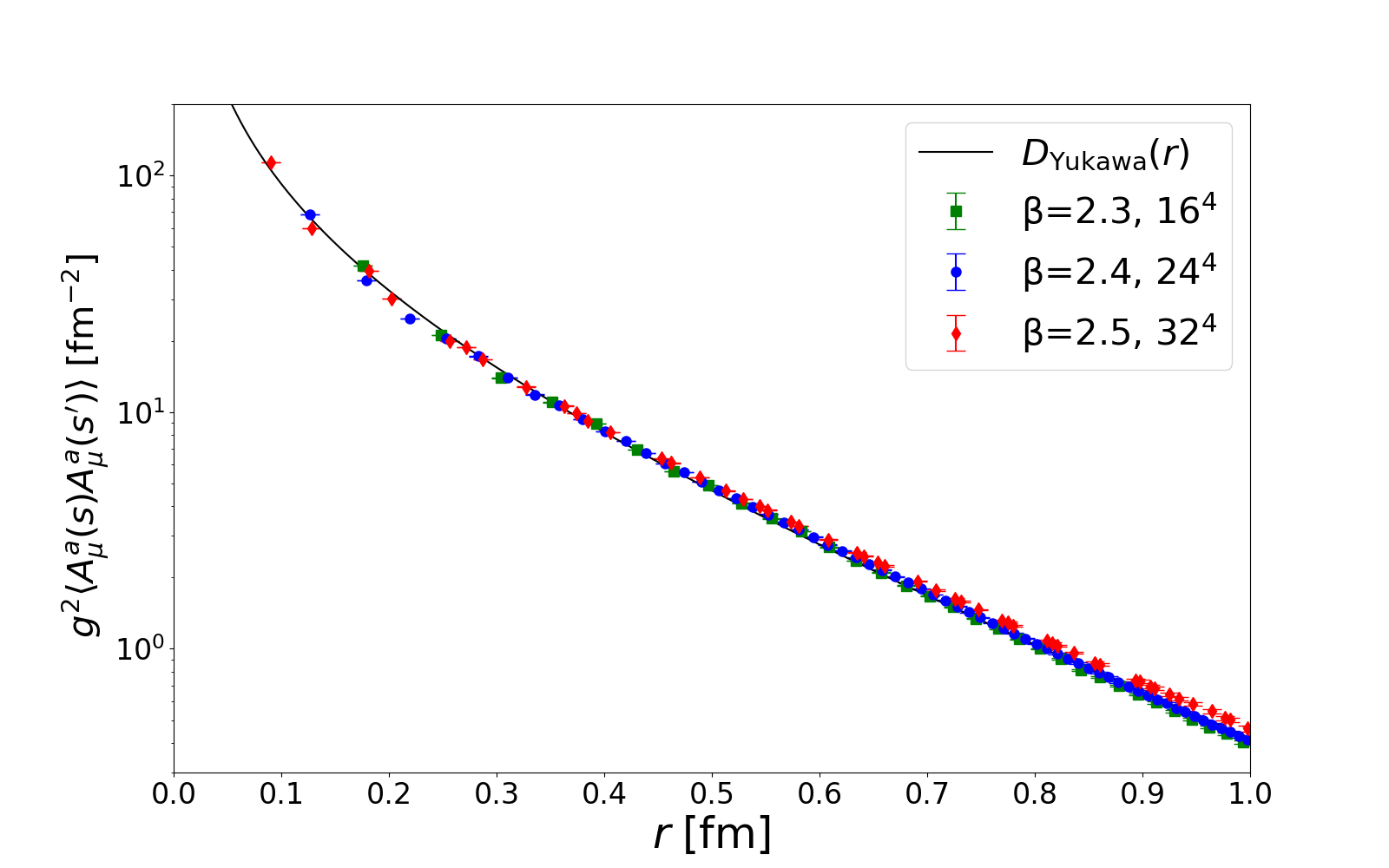}
    \caption{
Landau-gauge gluon propagator 
$g^2\langle A_\mu^a(s)A_\mu^a(s')\rangle$
plotted against $r\equiv |s-s'|$
in SU(3) (upper) and SU(2) (lower) lattice QCD. 
The curve is the best-fit 
Yukawa-type function $D_{\rm Yukawa}(r)$ 
in Eq.~(\ref{eq:Yukawa-type}). 
}
    \label{fig:prop}
\end{figure}
In both SU(2) and SU(3) QCD, 
the Landau-gauge gluon propagator $D(r)$ 
is well described with a Yukawa-type function 
in the wide region of $r = 0.1 - 1.0~{\rm fm}$, 
that is, 
\begin{eqnarray}
g^2 \langle A_\mu^a(s)A_\mu^a(s')\rangle
\simeq Am\frac{e^{-mr}}{r}\equiv D_{\rm Yukawa}(r),
\label{eq:Yukawa-type}
\end{eqnarray}
although 
the gluon propagator might take a different form  
in the far infrared region 
of $r \gg 1~{\rm fm}$
\cite{OS12}.

From the fit analysis with the Yukawa-type function as shown in Fig.\ref{fig:prop}, 
the gluonic mass parameter $m$ is estimated as 
$m\simeq 0.66~{\rm GeV}$ 
for SU(3) QCD \cite{ISI09}
and 
$m\simeq 0.676~{\rm GeV}$ 
for SU(2) QCD. 
The best-fit parameters are summarized in Table \ref{tab:Yukawa}.
[The definition of 
the coefficient $A$ is 
different from that in Ref.~\cite{ISI09} by 
$3(N_c^2-1)g^2$.]

\begin{center}
\begin{table}[htbp]
    \centering
    \begin{tabular}{ccc} 
    \hline
    \hline
    ~~~~~~~~SU$(N_c)$~~~~~~~~ & ~~~~~~~~ $m$~~~~~~~~ & ~~~~~~~~$A$~~~~~~~~ \\ 
    \hline
        SU(3) & 0.660(4) GeV & 5.87(2) \\
        SU(2) & 0.676(6) GeV & 3.78(2) \\
    \hline
    \hline
    \end{tabular}
    \caption{Best-fit
    parameters on the gluonic mass $m$ and the coefficient $A$ 
    in the Yukawa-type function $D_{\rm Yuakwa}(r)$
    in Eq.~(\ref{eq:Yukawa-type}) 
    for the Landau-gauge gluon propagator $g^2\langle A_\mu^a(s)A_\mu^a(s') \rangle$
    in SU(3) and SU(2) lattice QCD.}
    \label{tab:Yukawa}
\end{table}
\end{center}

Since the Yukawa-type gluon propagation 
is natural in the three-dimensional space-time instead of the four-dimensional one
\cite{ISI09}, this might relate to some dimensional reduction hidden in nonperturbative QCD \cite{TS24}.

Next, we examine 
every tensor component of $D_{\mu\nu}(s)$ 
using its Fourier image, 
\begin{eqnarray}
\tilde D_{\mu\nu}(p)\equiv \int d^4s~e^{-ip\cdot s}
D_{\mu\nu}(s).
\end{eqnarray}
In Euclidean Landau-gauge QCD,  
due to the $O(4)$ Lorentz symmetry, 
the tensor structure of 
$\tilde D_{\mu\nu}(p)$ is uniquely determined as 
\begin{eqnarray}
\tilde D_{\mu\nu}(p)
=(\delta_{\mu\nu}-\hat p_\mu \hat p_\nu ) \tilde D(p^2)
\label{eq:Landau-tensor}
\end{eqnarray}
with $p^2=p^\mu p^\mu$ and 
$\hat p^\mu \equiv p^\mu/\sqrt{p^2}$.

For the Yukawa-type propagator,
$D(r)=Ame^{-mr}/r$,
$\tilde D(p^2)$ is written by \cite{ISI09}
\begin{eqnarray}
\tilde D(p^2)=\frac13 \tilde D_{\mu\mu}(p)
=\frac{4\pi^2Am}{3~(p^2+m^2)^{3/2}},
\label{eq:Yukawa-mom}
\end{eqnarray}
and, as shown in Appendix A, 
we derive 
\begin{eqnarray}
D_{\mu\nu}(s-s')=\frac{Am^2}{3}\Big[F(mr)\delta_{\mu\nu}-G(mr)\hat s^R_\mu\hat s^R_\nu\Big]
\end{eqnarray}
with $r\equiv|s-s'|$, $\hat s^R \equiv (s-s')/r$ and 
\begin{eqnarray}
F(w)&\equiv &\frac{1}{w}\left[e^{-w}\left(1+\frac{1}{w}+\frac{2}{w^2}+\frac{2}{w^3}\right)-\frac{2}{w^3}\right],
\label{eq:F-fun} 
\\
G(w)&\equiv&\frac{1}{w}\left[e^{-w}\left(1+\frac{4}{w}+\frac{8}{w^2}+\frac{8}{w^3}\right)-\frac{8}{w^3}\right].
\label{eq:G-fun}
\end{eqnarray}

\section{Lattice QCD result for 
color-magnetic correlations} 

In this section, we study the color-magnetic correlations, $C_{\perp}(r)$ and $C_{\parallel}(r)$, in the Landau gauge 
using SU(2) and SU(3) lattice QCD 
at the quenched level.
Note again that 
all two-point field-strength correlations  
$g^2\langle G^a_{\mu\nu}(s)G^b_{\alpha\beta}(s') \rangle$ are expressed with these two correlations. 
For the lattice data analysis, 
we use rotational invariance 
via relabeling the space-time axis, 
as well as translational invariance, 
in Euclidean Landau-gauge QCD.

In the adopted $\beta$ region, 
$C_{\perp}(r)$ and 
$C_{\parallel}(r)$ at different $\beta$ values are found to be approximately single-valued functions 
of $r\equiv|s-s'|$ in lattice QCD.
(In Appendix B, we present the lattice QCD data on 
the color-magnetic correlations, $C_\perp$ and $C_\parallel$, for each $\beta$ in SU(3) and SU(2), respectively, to provide comprehensive information.)

\subsection{Perpendicular-type color-magnetic correlation}

\begin{figure}[htbp]
    \centering
\includegraphics[width=10cm]{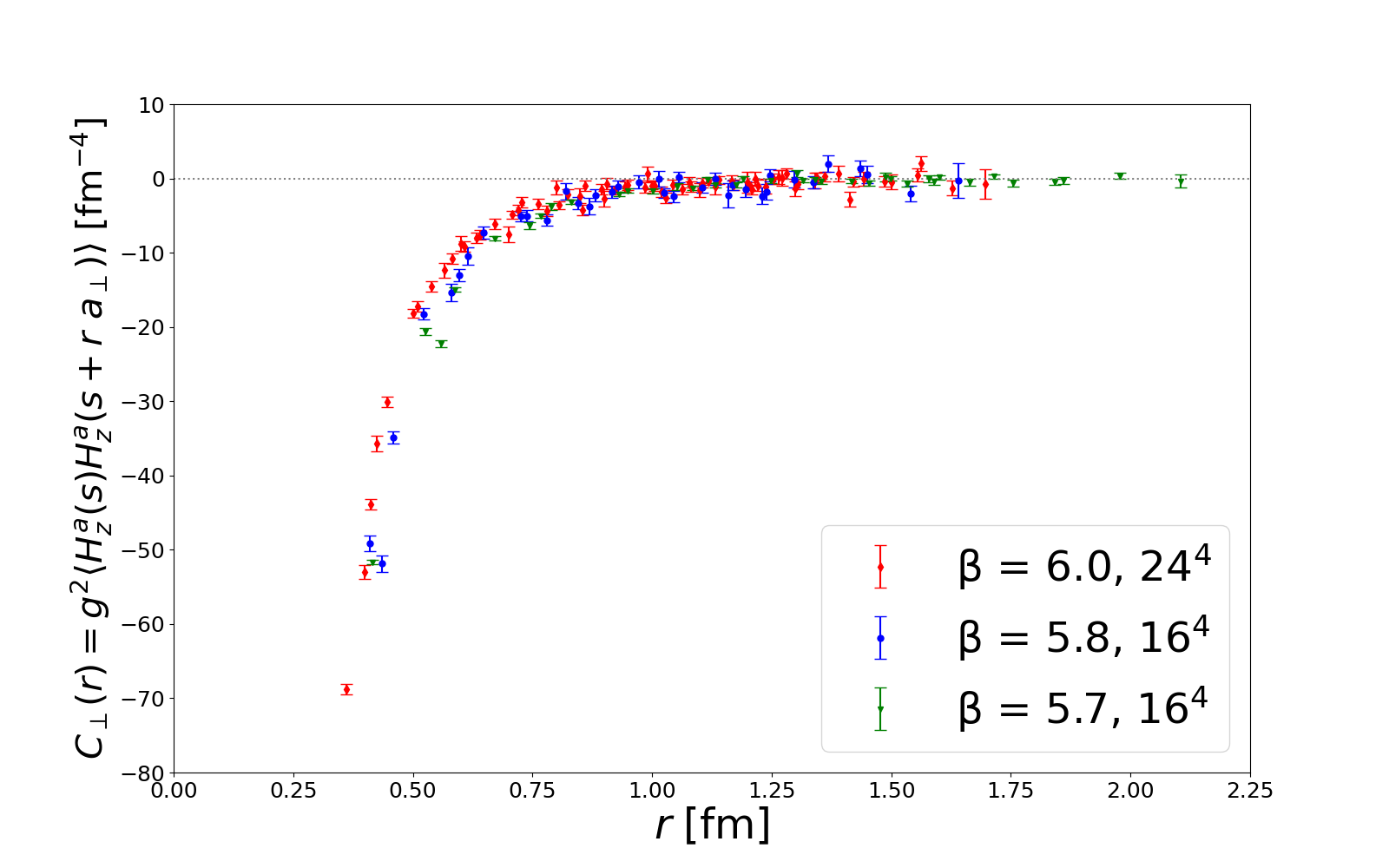}
\includegraphics[width=10cm]{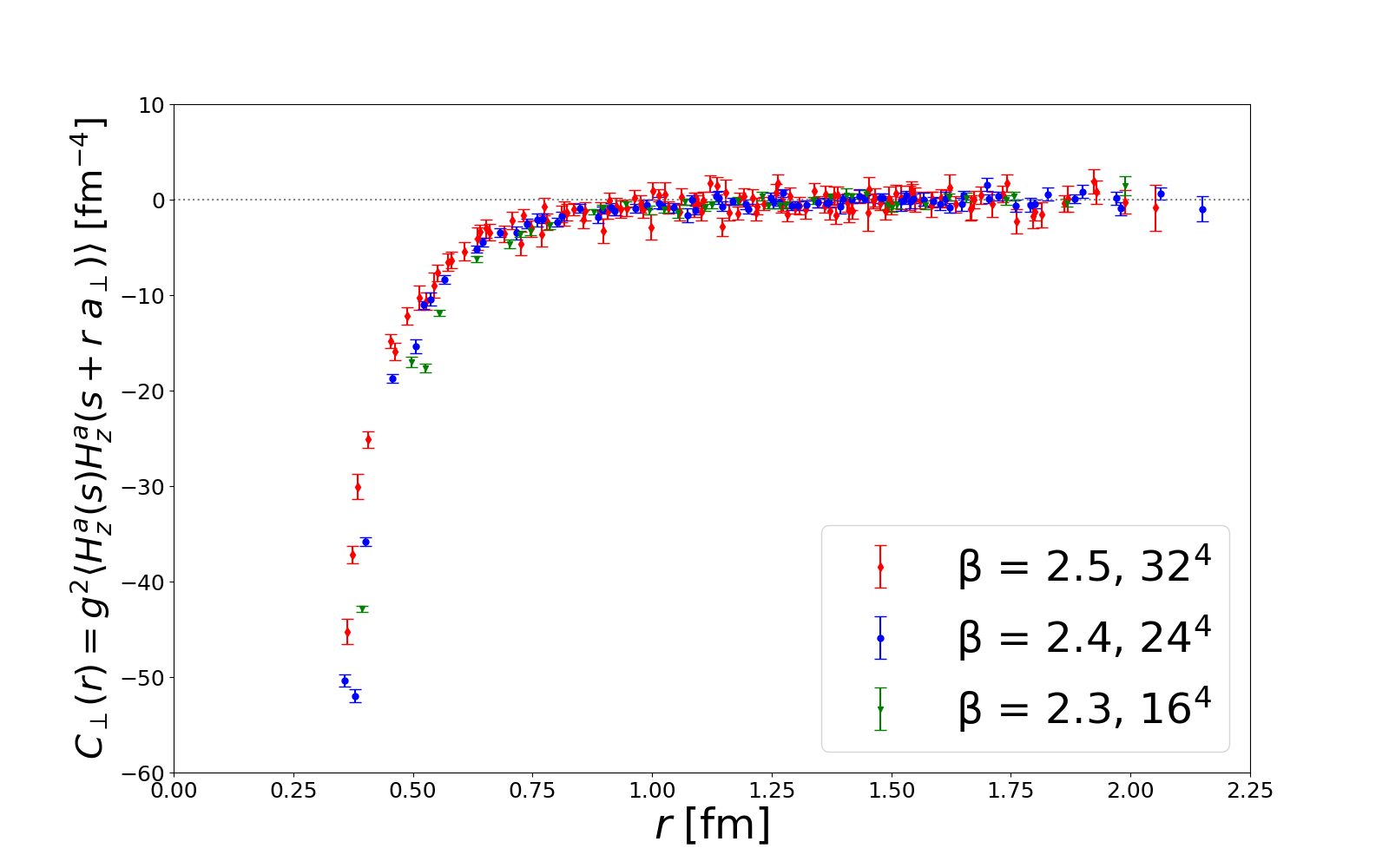}
    \caption{
    Perpendicular-type color-magnetic correlation
$C_{\perp}(r) \equiv g^2\langle H^a_z(s)H^a_z(s+r \hat \perp)\rangle$ 
    in the Landau gauge
    in SU(3) (upper) and SU(2) (lower) lattice QCD.
    $\hat \perp$ denotes 
    a unit vector on the $xy$-plane.
    }
    \label{fig:perp}
\end{figure}

First, we investigate the 
perpendicular-type color-magnetic correlation 
\begin{eqnarray}
C_{\perp}(r) \equiv g^2\langle H^a_z(s)H^a_z(s+r \hat \perp)\rangle 
\end{eqnarray} 
in the Landau gauge. 
Here, $\hat \perp$ is an arbitrary unit vector on the $xy$-plane.
Figure~\ref{fig:perp} shows the numerical result of $C_\perp(r)$
in SU(3) and SU(2) lattice QCD.

Curiously, the perpendicular-type color-magnetic correlation is {\it always negative}, 
$C_{\perp}(r)<0$, for all values of $r$, 
except for the same-point correlation at $r$ = 0, 
$C_{\perp}(0)=g^2 \langle H_z^a(s)H_z^a(s)\rangle$, 
which is trivially positive.
The lattice unit value of the same-point correlation $C_{\perp}(0)$ is as follows: 
In SU(3) lattice QCD, 
$C_{\perp}(0)=$ 7.20, 6.71, and 6.09  
at $\beta$=5.7, 5.8, and 6.0, respectively.
In SU(2) lattice QCD, 
$C_{\perp}(0)=$ 5.46, 4.63, and 4.09 
at $\beta$=2.3, 2.4, and 2.5, respectively. 

We note that an ``always negative" correlation is rare in physics, 
whereas ``always positive" and ``alternating" correlations 
have been observed in various areas of physics.

\begin{figure}[htbp]
    \centering
    \includegraphics[width=5cm]{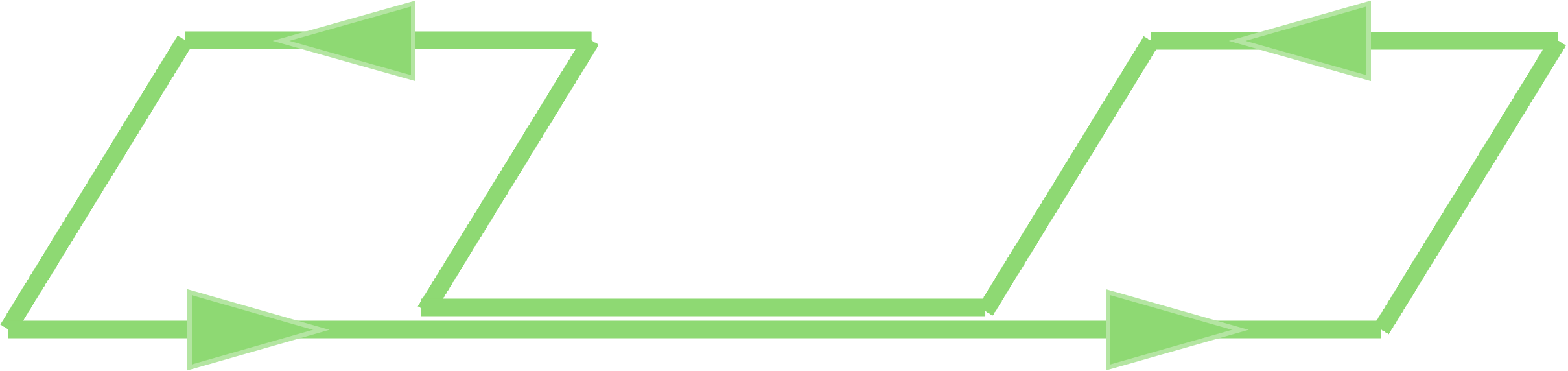}
    \caption{
    Example of the plaquette correlator 
    to extract the gauge-invariant
    field-strength correlation in lattice QCD. 
    }
    \label{fig:plaquette-correlator}
\end{figure}

One might suspect that the gauge fixing introduces an unphysical effect. 
However, we examined the gauge-invariant field-strength correlation extracted from the plaquette correlators, as shown in Fig.~\ref{fig:plaquette-correlator}, and obtained a similar result. In fact, the gauge-invariant correlation corresponding 
to the perpendicular-type color-magnetic correlation $C_\perp(r)$ is always negative, 
except for the same-point correlation $C_\perp(0)$, which is positive at $r$=0. 

Now, we explore the physical implications of $C_{\perp}(r)$ always being negative
except for $r$=0.
This seems to contradict the picture of continuously varying color-magnetic fields in the QCD vacuum. 
In fact, this lattice result suggests that the QCD vacuum differs significantly from homogeneous color-magnetic systems \cite{S77} or multi-vortex systems \cite{AO80}, even at small distances of the order of the lattice spacing $a= 0.1-0.2~{\rm fm}$. 

Instead, our result supports the picture that color-magnetic fields are highly stochastic in the QCD vacuum
\cite{D87,S88,DS88,GP92,EGM97,GMP97,GDSS02,EGM03,BBV98}, similar to the spaghetti vacuum \cite{NO79}. In other words, even if some color-magnetic domains exist, they would be tiny, and randomness dominates even at small distances of about $a= 0.1-0.2~{\rm fm}$.

\subsection{Parallel-type color-magnetic correlation}

Next, we investigate 
the parallel-type color-magnetic correlation 
\begin{eqnarray}
C_\parallel(r) \equiv g^2\langle H^a_z(s)H^a_z(s+r \hat \parallel)\rangle 
\end{eqnarray} 
in the Landau gauge.
Here, $\hat \parallel$ denotes 
an arbitrary unit vector on the $tz$-plane.

\begin{figure}[htbp]
    \centering
    \includegraphics[width=10cm]{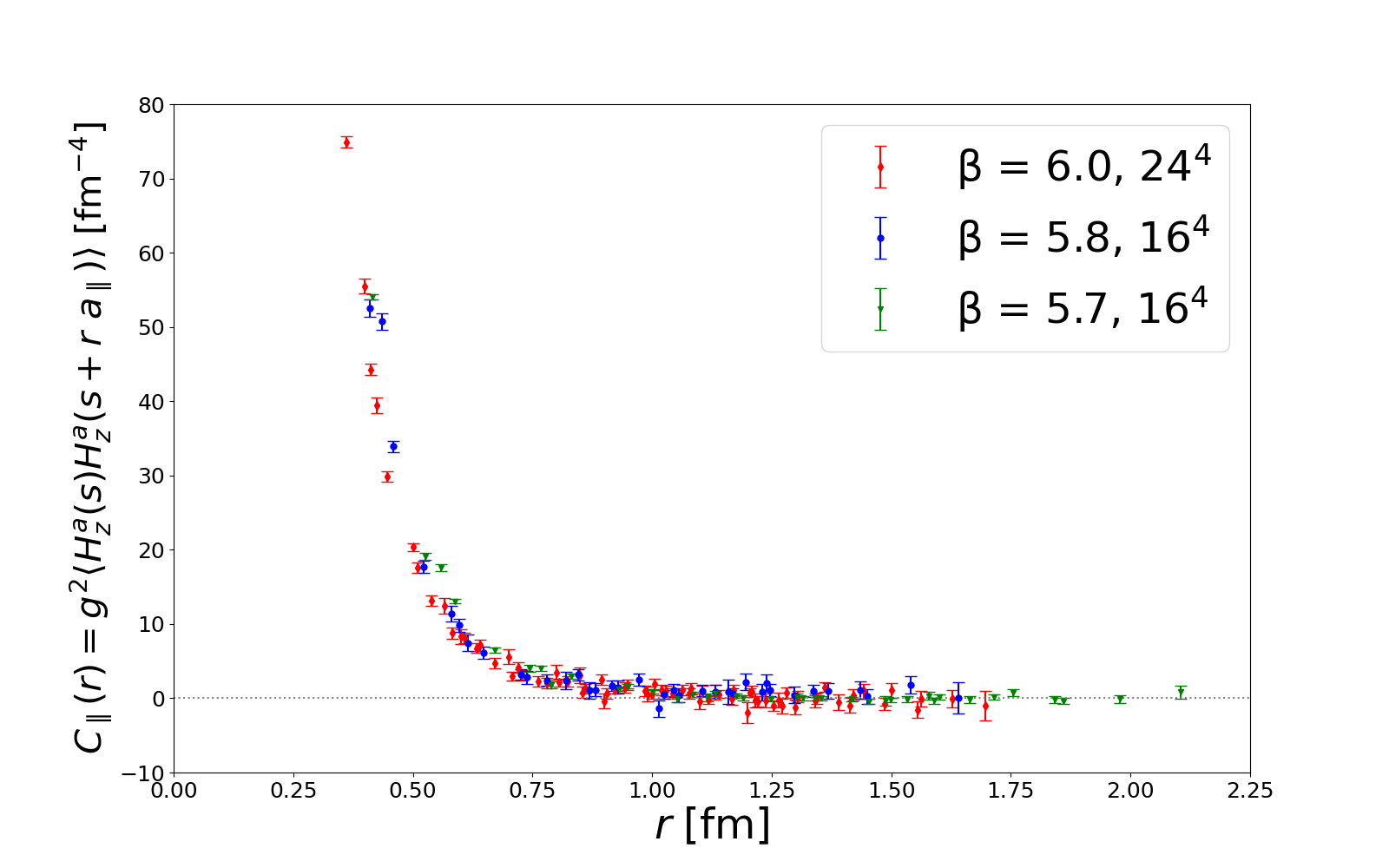}
    \includegraphics[width=10cm]{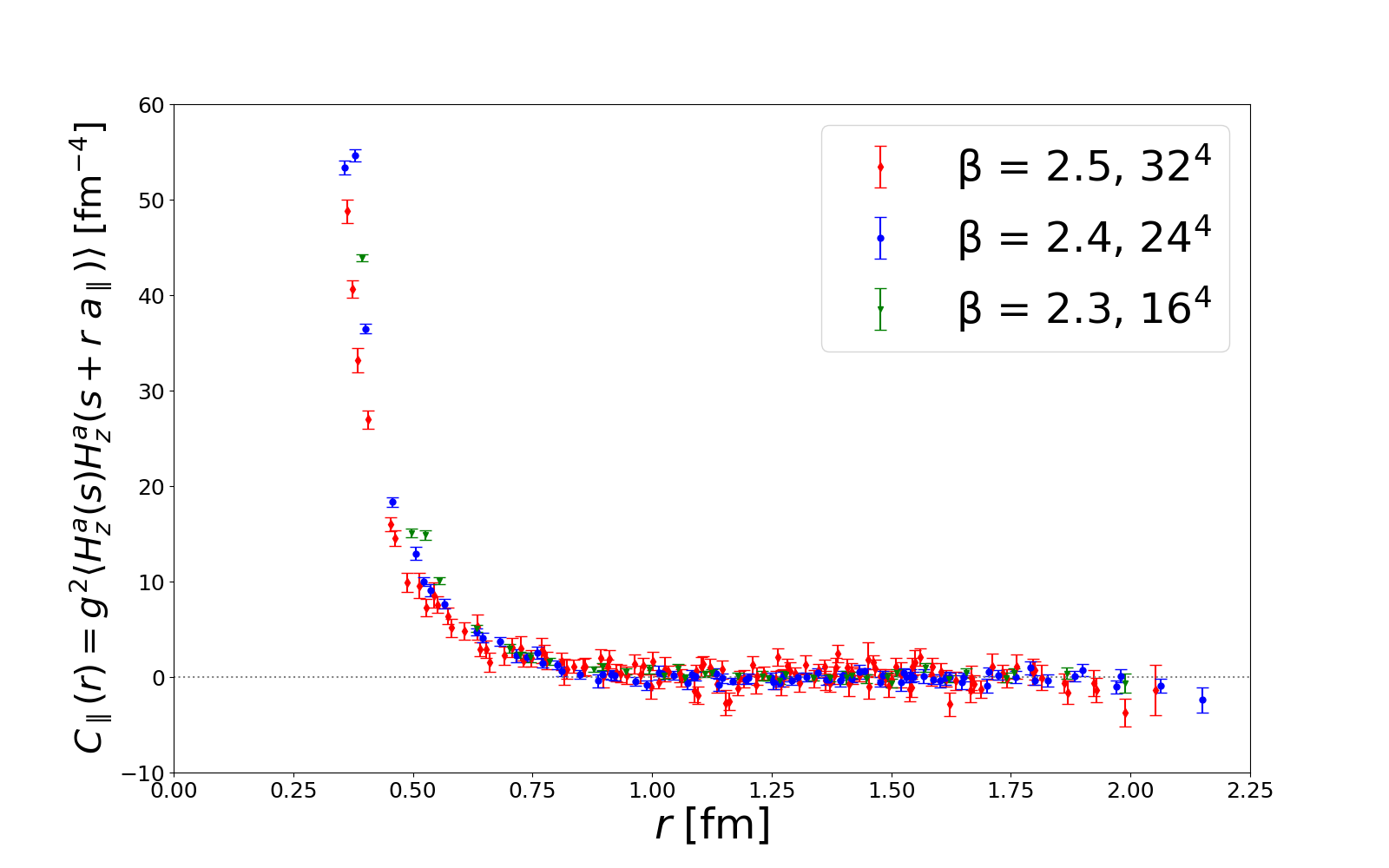}
    \caption{
    Parallel-type color-magnetic correlation
    $C_{\parallel}(r) \equiv g^2\langle H^a_z(s)H^a_z(s+r \hat \parallel)\rangle$
    in the Landau gauge
    in SU(3) (upper) and SU(2) (lower) lattice QCD.
    $\hat \parallel$ denotes a unit vector on the $zt$-plane.
    }
    \label{fig:para}
\end{figure}

Figure~\ref{fig:para} shows the 
numerical result of 
$C_{\parallel}(r)$ 
in SU(3) and SU(2) lattice QCD. 
The parallel-type color-magnetic  correlation is always positive, $C_{\parallel}(r) > 0$, 
for all distances $r$.

Unlike the case of $C_\perp(r)$, 
this positive correlation seems natural 
in continuum field theories.
In particular, in regular abelian gauge theories, such a positive correlation can be explained by magnetic flux conservation, although it is nontrivial in non-abelian cases. 

\begin{figure}[htbp]
    \centering
    \includegraphics[width=10cm]{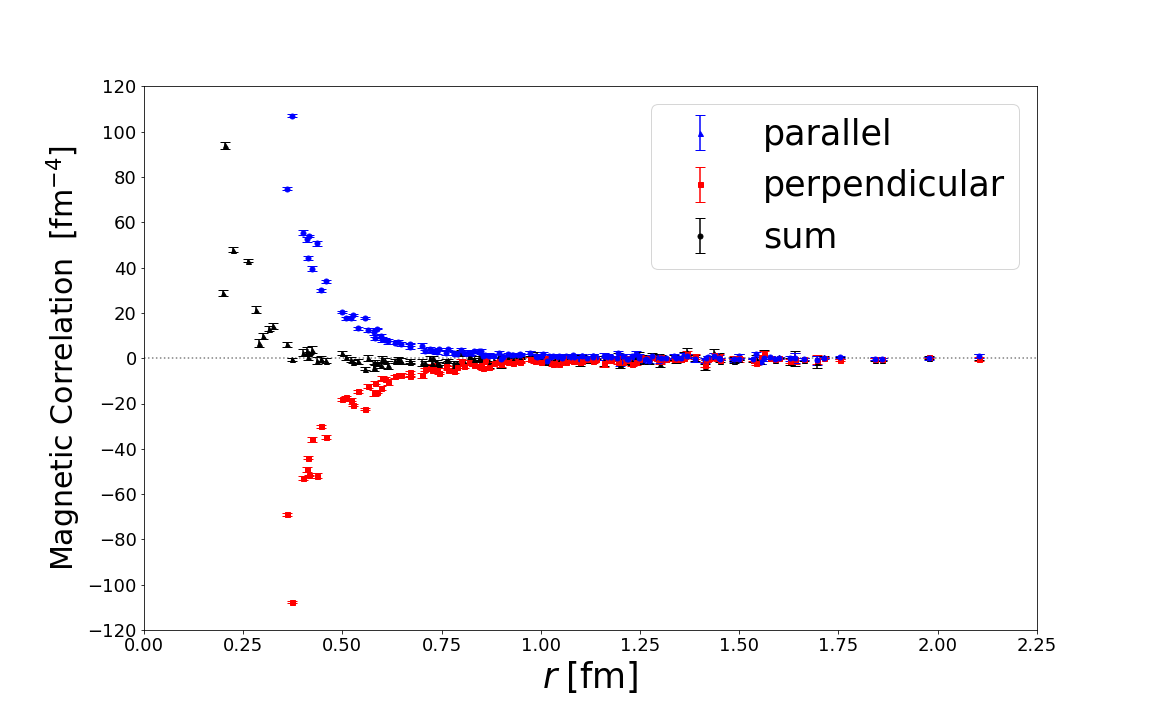}
    \includegraphics[width=10cm]{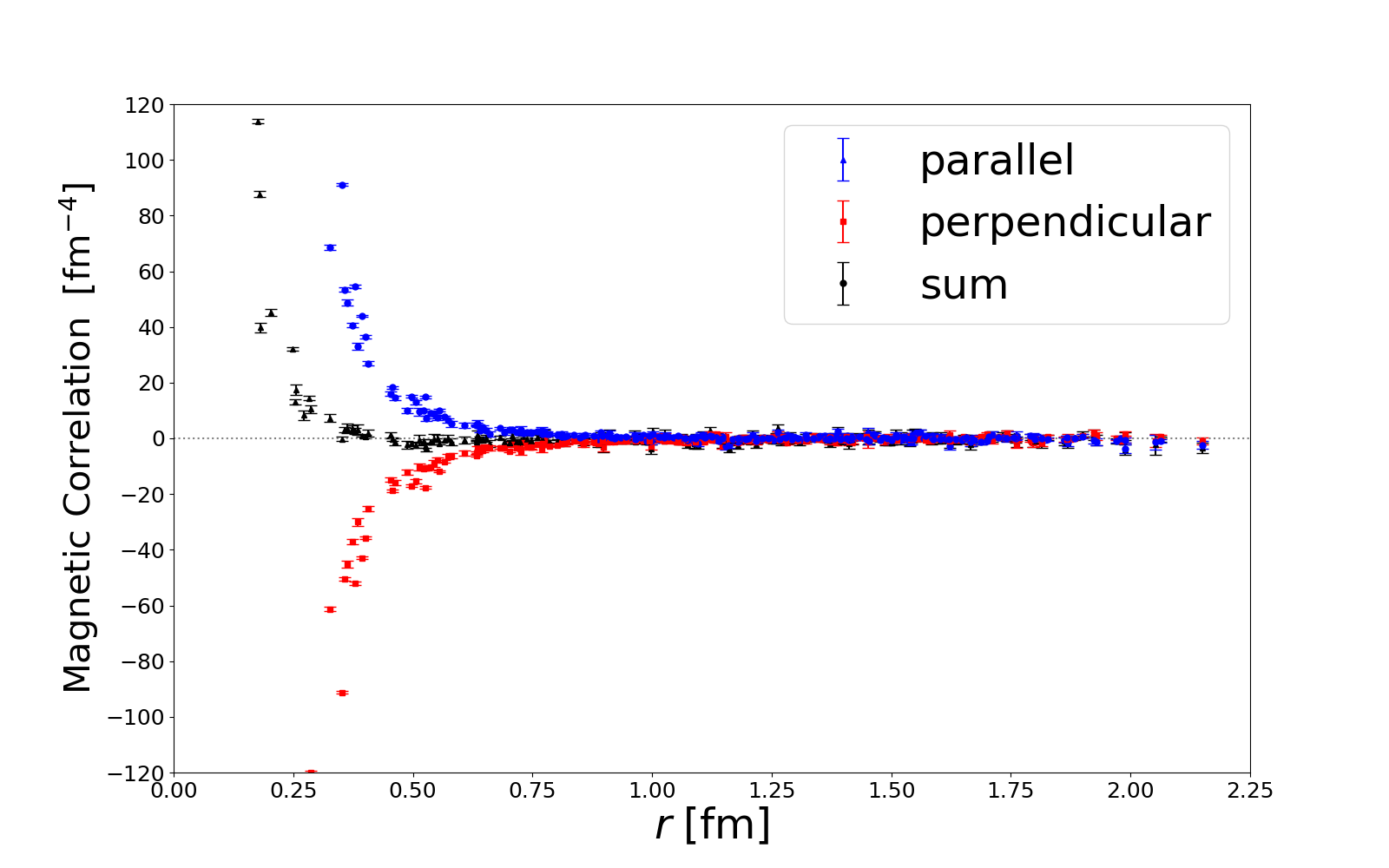}
    \caption{
    The color-magnetic correlation of 
    the parallel-type $C_{\parallel}(r)$, 
    the perpendicular-type $C_{\perp}(r)$,  
    and their sum $C_{\perp}(r)+C_{\parallel}(r)$
    in the Landau gauge
    in SU(3) (upper) and SU(2) (lower) lattice QCD.
    }
    \label{fig:perppara}
\end{figure}

Figure~\ref{fig:perppara} shows 
the color-magnetic correlation of 
the parallel-type $C_{\parallel}(r)$, 
the perpendicular-type $C_{\perp}(r)$,  and their sum $C_{\perp}(r)+C_{\parallel}(r)$
in the Landau gauge
in SU(3) and SU(2) lattice QCD.

For all values of $r$, 
the sum $C_{\perp}(r)+C_{\parallel}(r)$ decreases monotonically with $r$ and seems to be always positive,
\begin{eqnarray}
C_{\perp}(r)+C_{\parallel}(r)=
\frac16\sum_{\mu, \nu} g^2\langle G^a_{\mu\nu}(s)G^a_{\mu\nu}(s')\rangle 
>0
\end{eqnarray} 
with $r=|s-s'|$. 
For $r=0$, this sum leads to a positive value for the gauge-invariant gluon condensate, 
$g^2\langle G_{\mu\nu}^a G_{\mu\nu}^a \rangle > 0$.

In the infrared region of $r \gtrsim 0.4~{\rm fm}$, 
the parallel-type color-magnetic correlation $C_{\parallel}(r)$
strongly cancels with 
the perpendicular-type correlation 
$C_{\perp}(r)$
as 
\begin{eqnarray}
C_{\parallel}(r) \simeq - C_{\perp}(r), 
\end{eqnarray}
which physically means 
\begin{eqnarray}
\langle H^a_z(s)H^a_z(s+r\hat z)\rangle 
&\simeq& 
-\langle H^a_z(s)H^a_z(s+r\hat x)\rangle,
\end{eqnarray}
and such a relation leads to a strong cancellation 
for the sum of Landau-gauge field-strength correlations,
\begin{eqnarray}
\sum_{\mu, \nu} g^2\langle G^a_{\mu\nu}(s)G^a_{\mu\nu}(s')\rangle 
=6~[C_{\perp}(r)+C_{\parallel}(r)]
\simeq 0,~~~
\end{eqnarray}
in the four-dimensional Euclidean space-time.

In Appendix C, we also consider 
the correlation of 
the SU(3) ``plaquette field strength'' ${\cal G}_{\mu\nu}$
defined in Eq.~(\ref{eq:P-elemag}). 
Regarding the lattice result, 
the field-strength correlation 
defined by ${\cal G}_{\mu\nu}$ 
is similar to that defined by $G_{\mu\nu}$. 

\section{Decomposition and Discussion}

In this section, we try to analyze 
the lattice QCD result 
of the color-magnetic correlation 
in the Landau gauge, particularly considering the origin of 
the negative correlation of $C_{\perp}(r)<0$.

\subsection{Decomposition of the field-strength correlation in terms of the gluon field}


In terms of the number of 
the gluon field $A_\mu$, 
we decompose the field-strength correlation 
$\langle G_{\mu\nu}^a (s) G_{\alpha\beta}^a (s') \rangle$
into three parts, i.e.,  
quadratic, cubic and quartic terms: 
\begin{eqnarray}
\langle G_{\mu\nu}^a (s) G_{\alpha\beta}^a (s') \rangle
&=&\langle G_{\mu\nu}^a (s) G_{\alpha\beta}^a (s') \rangle_{\rm quad} \cr &+& 
 \langle G_{\mu\nu}^a (s) G_{\alpha\beta}^a (s') \rangle_{\rm cubic} \cr
&+& \langle G_{\mu\nu}^a (s) G_{\alpha\beta}^a (s') \rangle_{\rm quartic}. 
\end{eqnarray} 
The quadratic, cubic and quartic terms are defined by
\begin{eqnarray}
&&\langle G_{\mu\nu}^a (s) G_{\alpha\beta}^a (s') \rangle_{\rm quad}\cr
&\equiv&  \langle  
(\partial_\mu A^a_\nu-\partial_\nu A^a_\mu) (s) 
(\partial_\alpha A^a_\beta-\partial_\beta A^a_\alpha) (s') \rangle,
\cr \cr
&&\langle G_{\mu\nu}^a (s) G_{\alpha\beta}^a (s') \rangle_{\rm cubic}\cr
&\equiv& 2ig\langle {\rm Tr}\{ 
(\partial_\mu A_\nu-\partial_\nu A_\mu) (s) 
[A_\alpha, A_\beta] (s') \}\rangle \cr
&+&
2ig\langle {\rm Tr}\{
[A_\mu, A_\nu] (s)
(\partial_\alpha A_\beta-\partial_\beta A_\alpha) (s') 
 \}\rangle, 
\cr \cr
&&\langle G_{\mu\nu}^a (s) G_{\alpha\beta}^a (s') \rangle_{\rm quartic}\cr
&\equiv& -2g^2\langle {\rm Tr}\{ 
[A_\mu, A_\nu] (s) [A_\alpha, A_\beta] (s') \}\rangle,
\end{eqnarray}
where the factor of two comes from 
${\rm Tr}(T^aT^b)=\frac 12\delta^{ab}$.

Among the three terms, the quadratic term can be directly expressed with the gluon propagator (\ref{eq:propagator}), i.e., 
$D_{\mu\nu}(s-s')
\equiv
g^2 \langle A_\mu^a(s)A_\nu^a(s') \rangle$, as
\begin{eqnarray}
&&g^2\langle 
G_{\mu\nu}^a (s) G_{\alpha\beta}^a (s') \rangle_{\rm quad} \cr
&=&
\partial_\mu^s \partial_\alpha^{s'} 
D_{\nu\beta}(s-s') 
-\partial_\mu^s \partial_\beta^{s'} 
D_{\nu\alpha}(s-s') \cr
&-&\partial_\nu^s \partial_\alpha^{s'} 
D_{\mu\beta}(s-s') 
+\partial_\nu^s \partial_\beta^{s'} 
D_{\mu\alpha}(s-s').
\end{eqnarray}
In the Laudau gauge, due to the Lorentz symmetry, this quantity can be expressed using the scalar combination of the gluon propagator $D(r) \equiv g^2 \langle A_\mu^a(s)A_\mu^a(s') \rangle$, which is a single-valued function of the four-dimensional Euclidean distance $r \equiv |s-s'|$.

\subsection{Decomposition of perpendicular-type color-magnetic correlation in the Landau gauge}

In terms of the gluon field $A_\mu$, 
the color-magnetic correlation 
$\langle H_z^a (s) H_z^a (s') \rangle$
is decomposed as 
\begin{eqnarray}
&&\langle H_z^a (s) H_z^a (s') \rangle \cr
&=& \langle  
(\partial_x A^a_y-\partial_y A^a_x) (s) 
(\partial_x A^a_y-\partial_y A^a_x) (s') \rangle \cr
&+&4ig\langle {\rm Tr}\{ 
(\partial_x A_y-\partial_y A_x) (s) 
[A_x, A_y] (s') \}\rangle
\cr
&-&2g^2\langle {\rm Tr}\{ 
[A_x, A_y] (s) [A_x, A_y] (s')\} \rangle,
\end{eqnarray}
and the quadratic term of $A_\mu$ can be described by the gluon propagator 
$D_{\mu\nu}(s-s')$, or more specifically by $D(r)$ in the Landau gauge. 

In the case of the Yukawa-type gluon propagator 
(\ref{eq:Yukawa-type}),  
we calculate the quadratic term 
in the color-magnetic correlation
at the same time point of 
$s-s'=(x,y,z,0)$ as 
\begin{widetext}
\begin{eqnarray}
g^2 \langle H_z^a(s)H_z^a(s')\rangle_{\rm quad} &\equiv& g^2\langle (\partial_1 A_2^a-\partial_2 A_1^a)(s)(\partial_1 A_2^a-\partial_2 A_1^a)(s')\rangle 
=g^2 \epsilon_{ij3}\epsilon_{kl3}\partial_i^s \partial_k^{s'}
\langle A_j^a(s) A_l^a(s')\rangle 
\cr
&=&-\epsilon_{ij3}\epsilon_{kl3}\partial_i^s \partial_k^{s}
D_{jl}(s-s') \cr
&=&
-\frac{Am^2}{3}\left[\frac{1}{\rho}\frac{\partial}{\partial \rho}\rho \frac{\partial}{\partial \rho}F(mr)+\frac{1}{r}\frac{
d}{dr} G(mr)
+\frac{z^2}{r^2}\left\{\frac{2}{r^2}G(mr)-\frac{1}{r}\frac{d}{dr}G(mr)\right\}\right]
\label{eq:Yukawa-derived}
\end{eqnarray}    
\end{widetext}
with 
the cylindrical radial coordinate 
$\rho \equiv 
(x^2+y^2)^{1/2}$
and $r\equiv |s-s'|=(\rho^2+z^2)^{1/2}$. 
(See Appendix D.)

For the Yukawa-type gluon propagator, 
the quadratic term in the perpendicular-type color-magnetic correlation 
$C_\perp(r)$ ($\hat \perp \in xy\hbox{-plane}$) is always negative, 
\begin{eqnarray}
&&g^2\langle H_z^a(s) H_z^a(s+r \hat \perp)\rangle_{\rm quad}\cr
&=&g^2\langle  
(\partial_x A^{a}_y-\partial_y A^{a}_x) (s) 
(\partial_x A^{a}_y-\partial_y A^{a}_x) (s+r \hat \perp) \rangle \cr
&=&-\frac{Am^4}{3}~\frac{1}{mr}\left(1+\frac{1}{mr}+\frac{1}{m^2r^2}\right)e^{-mr} < 0, 
\label{eq:Yukawa-perp}
\end{eqnarray}
and the quadratic term in the parallel-type color-magnetic
correlation $C_\parallel(r)$ ($\hat \parallel \in zt\hbox{-plane}$) is always positive, 
\begin{eqnarray}
&&g^2 \langle H_z^a(s)H_z^a(s+r \hat \parallel)\rangle_{\rm quad}  \cr
&=&g^2\langle  
(\partial_x A^{a}_y-\partial_y A^{a}_x) (s) 
(\partial_x A^{a}_y-\partial_y A^{a}_x) (s+r \hat \parallel) \rangle \cr
&=&\frac{2Am^4}{3}\frac{1}{m^2r^2}
\left(1+\frac{1}{mr} \right)e^{-mr} >0.
\end{eqnarray} 

Then, if the quadratic term is dominant, 
the negative behavior of the perpendicular-type color-magnetic correlation, 
$C_{\perp}(r) < 0$, could be explained by the Yukawa-type gluon propagator. 
However, the real situation is not so simple.

\begin{figure}[htbp]
    \centering
    \includegraphics[width=10cm]{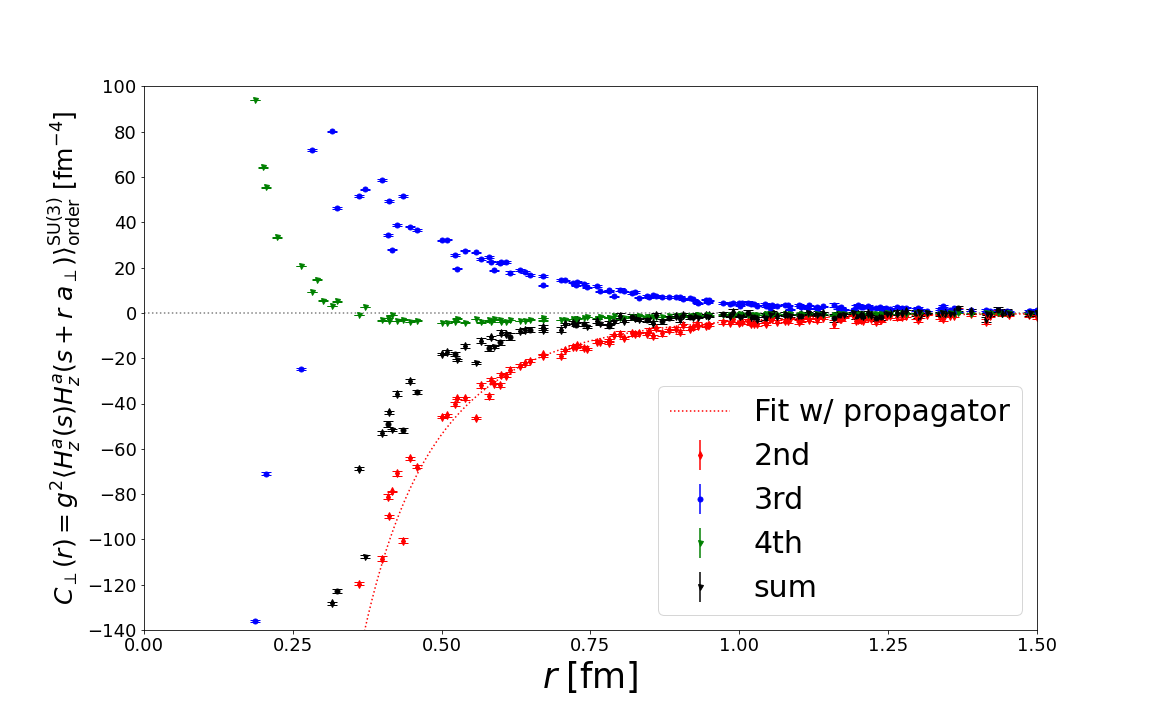}
    \includegraphics[width=10cm]{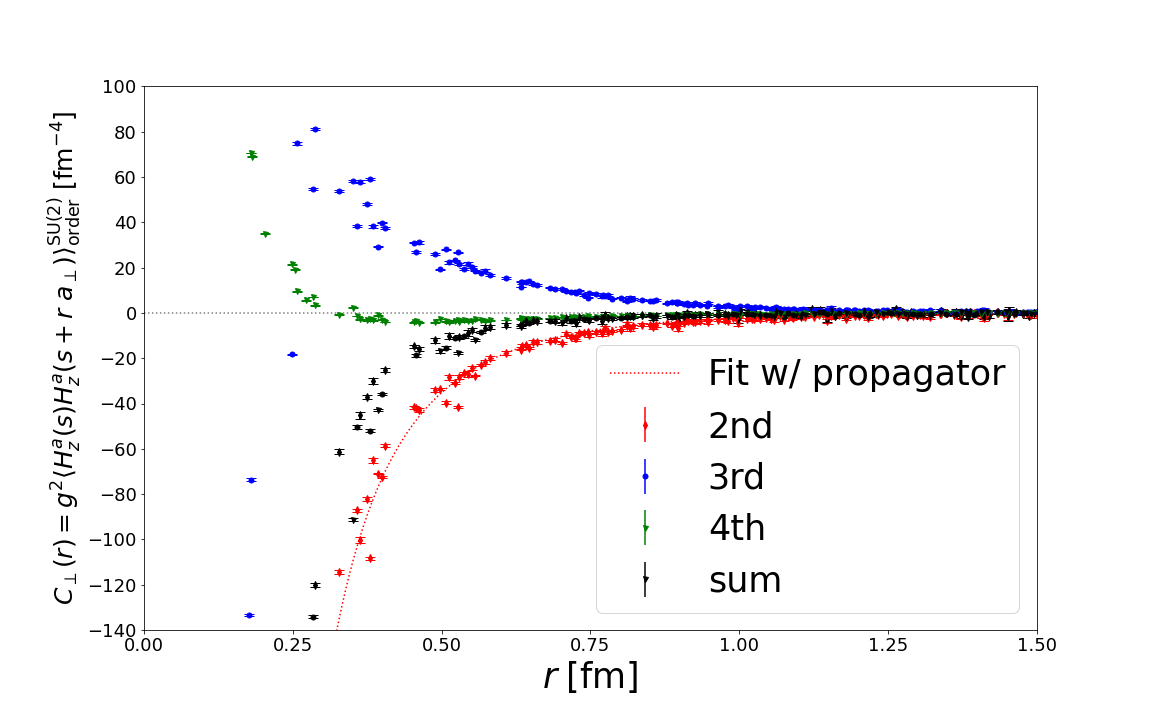}
    \caption{
Each contribution of 
the quadratic (red), cubic (blue) and quartic (green)
terms of the perpendicular-type color-magnetic correlation $C_{\perp}(r)$ (black)
in the Landau gauge in SU(3) (upper) and SU(2) (lower) lattice QCD. The red dotted line denotes the curve of Eq.~(\ref{eq:Yukawa-perp}), derived from the Yukawa-type propagator 
(\ref{eq:Yukawa-type}).
    }
    \label{fig:decomposition}
\end{figure}

Figure~\ref{fig:decomposition} 
shows individual contributions of 
the quadratic, cubic and quartic 
terms in the perpendicular-type color-magnetic correlation $C_{\perp}(r)$ in SU(3) and SU(2) lattice QCD.

The quadratic term is always negative,
as was demonstrated with the Yukawa-type gluon propagator in the Landau gauge. 
The quadratic term seems to be 
quantitatively reproduced by the Yukawa-type gluon propagator, 
as shown in Fig.~\ref{fig:decomposition}.

The cubic term is positive for large $r$ and negative for small $r$. 
In the infrared region, 
the magnitude of the cubic term  is similar to that of the quadratic term. 
Consequently,
the positive cubic term 
tends to cancel with 
the negative quadratic term, 
resulting in a small value of $C_{\perp}(r)$, 
in the infrared region.

Since the cubic term of the gauge field is unique to non-abelian gauge theories, its significant contribution in the QCD vacuum indicates the distinction between QCD and abelian gauge theories.

%

Compared to the quadratic and cubic terms, the quartic term contributes relatively little as an absolute value.
In the infrared region, however, the contribution of 
the quartic term is similar in magnitude to the sum of the contributions of the quadratic and cubic terms,  
because they cancel each other.

\section{Summary and Conclusion}

Using SU(2) and SU(3) quenched lattice QCD, we have studied the field-strength correlation 
$g^2 \langle G_{\mu\nu}^a(s)G^b_{\alpha\beta}(s') \rangle$, as well as the gluon propagator, in the Landau gauge.

To begin with, we have investigated the Landau-gauge gluon propagator $D(r)\equiv g^2 \langle A_\mu^a (s)A_\mu^a(s') \rangle$ and have found that it  
is well described by the Yukawa-type function 
$e^{-mr}/r$ with $r\equiv |s-s'|$ 
for $r=0.1-1.0~{\rm fm}$
in both SU(2) and SU(3) lattice QCD.

We have investigated 
the perpendicular-type color-magnetic correlation,   
$C_{\perp}(r) \equiv g^2\langle H_z^a(s)H_z^a(s + r \hat \perp)) \rangle$ ($\hat \perp \in  xy\hbox{-plane}$), 
and the parallel-type one,   
$C_{\parallel}(r) \equiv g^2 \langle H_z^a(s)H_z^a(s + r \hat \parallel) \rangle$ ($\hat \parallel \in zt\hbox{-plane}$). 
These two quantities reproduce 
all the correlation of 
$g^2\langle G^a_{\mu\nu}(s)G^b_{\alpha\beta}(s')\rangle$, 
due to the Lorentz and global SU($N_c$) color symmetries in the Landau gauge.

Curiously, we have found that the perpendicular-type color-magnetic correlation 
$C_{\perp}(r)$ is always negative for 
all values of $r$, 
except for the same-point correlation. 
In contrast, the parallel-type color-magnetic correlation $C_{\parallel}(r)$
is always positive. 
In the infrared region of $r\gtrsim 0.4~{\rm fm}$, 
$C_{\perp}(r)$ and $C_{\parallel}(r)$ 
strongly cancel each other, 
which leads to a significant  cancellation in the sum of the field-strength correlations as 
$\sum_{\mu, \nu} g^2\langle G^a_{\mu\nu}(s)G^a_{\mu\nu}(s')\rangle 
\propto C_{\perp}(|s-s'|)+ C_{\parallel}(|s-s'|) 
\simeq 0$.

Finally, we have decomposed  the field-strength correlation 
into quadratic, cubic and quartic terms of the gluon field $A_\mu$ in the Landau gauge.
For the perpendicular-type color-magnetic correlation $C_{\perp}(r)$,  
the quadratic term is always negative, which is explained 
by the Yukawa-type gluon propagator. 
The quartic term gives a relatively small contribution.
In the infrared region, 
the cubic term is positive and tends to cancel with the quadratic term, 
resulting in a small value of $C_{\perp}(r)$.

We have considered the physical implications of 
the negativity of the perpendicular-type color-magnetic correlation,
$C_{\perp}(r)<0$, which seems to contradict the picture of continuously varying color-magnetic fields.
This result suggests that the QCD vacuum differs significantly from homogeneous color-magnetic systems \cite{S77} or multi-vortex systems \cite{AO80}, even at small distances of the order of the lattice spacing $a= 0.1-0.2~{\rm fm}$. 

In abelian gauge theories, 
the negativity of the perpendicular-type magnetic correlation could be explained 
by magnetic flux conservation, although the ordinary QED vacuum is trivial.
For example, center projected QCD in the maximal center gauge \cite{DMGO97} would yield 
a negative value for the perpendicular-type magnetic correlation. 
However, in non-abelian gauge theories, such a simple explanation 
is not applied, and the negative color-magnetic correlation of 
$C_{\perp}(r) < 0 $ remains a puzzle in QCD. 

In this context, it is interesting to study magnetic correlations 
in Abelian-projected QCD in the maximally Abelian gauge, instead of the Landau gauge. 
In Abelian-projected QCD, 
all magnetic correlations can be expressed using the diagonal-gluon propagator \cite{AS99,BCGMP03,GS13}.
Despite the abelian gauge theory, 
the perpendicular-type magnetic correlation is nontrivial 
because magnetic flux conservation or the abelian Bianchi identity is violated by the appearance of magnetic monopoles
\cite{tH81,IS00}.

Finally, we discuss the physical consequences of the fact that $C_{\perp}(r)$ is always negative 
except for $r=0$. 
This indicates that the QCD vacuum differs significantly from homogeneous color-magnetic \cite{S77} or multi-vortex systems \cite{AO80},
even at distances of the order of the lattice spacing $a = 0.1-0.2~{\rm fm}$. 
Instead, our result supports the picture that color-magnetic fields are highly stochastic in the QCD vacuum
\cite{NO79,AO80,AKL82,D87,S88,DS88,GP92,EGM97,GMP97,GDSS02,EGM03,BBV98}, such as the spaghetti vacuum \cite{NO79}. In other words, even if some color-magnetic domains exist, they would be tiny, and randomness dominates even at the small distance scales.

In the context of stochastic magnetic fields, there is an interesting physical case of a spin system.
In 1979, Parisi and Sourlas demonstrated that a three-dimensional spin system in the presence of Gaussian random magnetic fields can be mathematically described as a one-dimensional spin system in the absence of magnetic fields \cite{PS79}.
In fact, dimensional reduction occurs for the spin system in a stochastic magnetic field.

As an interesting possibility, the random color-magnetic fields 
at the infrared scale might lead to some dimensional reduction effects for the QCD vacuum \cite{ISI09,TS24}, like the Parisi-Sourlas mechanism \cite{PS79}.
Indeed, the Yukawa-type gluon propagator could be interpreted as 
a sign of dimensional reduction of the QCD vacuum \cite{ISI09}. 
It is interesting to explore
the possibility that infrared dimensional reduction occurs
through the random color-magnetic fields in the QCD vacuum.

\section*{Acknowledgment}

We thank Professor George Savvidy for his useful comment on the stochastic nature of the vacuum fields in the Yang-Mills theory. 
We thank Professor Jeff Greensite for his useful comment and discussion on the case of abelian gauge theories. 
A.T. is supported by SPRING Program at Kyoto University.
K.T. is supported by JSPS Research Fellowship for Young Scientists. 
H.S. was supported in part 
by a Grants-in-Aid for Scientific Research [19K03869] from JSPS. 
The lattice QCD calculations have been performed by SQUID at RCNP at Osaka University.

\appendix

\section{Landau-gauge gluon propagator and Yukawa-type function}

In Appendix A, we investigate 
the Landau-gauge gluon propagator $D_{\mu\nu}(s-s')\equiv g^2\langle A^a_\mu(s)A^a_{\nu}(s')\rangle$
and the Yukawa-type propagator (\ref{eq:Yukawa-type}), which gives an approximation of 
$D_{\mu\mu}(r)$ for $r=0.1-1.0~{\rm fm}$.

\subsection{General formulae on Landau-gauge propagator}

First, we present the general formalism 
of the Landau-gauge propagator of gauge fields $A_\mu(s)$ in most gauge theories, including QCD and QED, in four-dimensional Euclidean space-time.

Using the inverse Fourier transform
of Eq.~(\ref{eq:Landau-tensor}), 
the coordinate-space propagator $D_{\mu\nu}(s)$ is generally expressed as 
\begin{eqnarray}
D_{\mu\nu}(s)
&=&\int \frac{d^4p}{(2\pi)^4}\tilde D(p^2) (\delta_{\mu\nu}-\hat p_\mu \hat p_\nu)e^
{ip\cdot s} \cr 
&=&
f(r)\delta_{\mu\nu}-g(r)
\hat s_\mu \hat s_\nu
\label{eq:Landau-gluon}
\end{eqnarray}
with the four-dimensional distance $r=|s|$
and $\hat s \equiv s/r$.
From the relations 
\begin{eqnarray}
D_{\mu\mu}(s) &=& 4f(r)-g(r),
\cr
\hat s_\mu \hat s_\nu D_{\mu\nu}(s) &=& f(r)-g(r),
\end{eqnarray}
one finds 
\begin{eqnarray}
f(r) &=& \frac13 [D_{\mu\mu}(s)- \hat s_\mu \hat s_\nu D_{\mu\nu}(s)],
\label{eq:f-Yukawa}
\\
g(r) &=&\frac13 [D_{\mu\mu}(s)- 4\hat s_\mu \hat s_\nu D_{\mu\nu}(s)].
\label{eq:g-Yukawa}
\end{eqnarray}
Using the Poisson integral representation,
\begin{eqnarray}
J_n(z)=\frac{(z/2)^n}{\sqrt{\pi}\Gamma(n+\frac12)}\int_0^\pi d\theta \cos(z\cos\theta) \sin^{2n}\theta, ~~~~
\end{eqnarray}
we obtain 
\begin{eqnarray}
D_{\mu\mu}(s)
&=&3\int \frac{d^4p}{(2\pi)^4}\tilde D(p^2) e^
{ip\cdot s} \cr
&=&\frac{3}{4\pi^3}\int_0^\infty 
dp p^3\tilde D(p^2) \int_0^\pi 
d\theta~
\sin^2\theta~e^{ipr \cos \theta}
\cr 
&=& \frac{3}{4\pi^2r}
\int_0^\infty dp p^2\tilde D(p^2) J_1(pr),
\end{eqnarray}
\begin{eqnarray}
\hat s_\mu \hat s_\nu D_{\mu\nu}(s)
&=&\int \frac{d^4p}{(2\pi)^4}[1-(\hat p\cdot \hat s)^2]\tilde D(p^2) e^
{ip\cdot s} \cr
&=&\frac{1}{4\pi^3}\int_0^\infty 
dp p^3\tilde D(p^2) \int_0^\pi 
d\theta~
\sin^4\theta~e^{ipr \cos \theta}
\cr 
&=& \frac{3}{4\pi^2r^2}
\int_0^\infty dp p\tilde D(p^2) J_2(pr).
\end{eqnarray}

\subsection{Yukawa-type propagator}

Next, we investigate the Landau-gauge propagator $D_{\mu\nu}(s)$ corresponding to 
the Yukawa-type propagator (\ref{eq:Yukawa-type}).
For the Yukawa-type propagator, the expression of $\tilde D(p^2)$ is given by Eq.~(\ref{eq:Yukawa-mom}) \cite{ISI09}, 
and we obtain 
\begin{eqnarray}
&& D_{\mu\mu}(s)
= \frac{3}{4\pi^2r}
\int_0^\infty dp p^2\tilde D(p^2) J_1(pr) \cr
&=& \frac{Am}{r}
\int_0^\infty dp~ \frac{p^2}{(p^2+m^2)^{3/2}} J_1(pr) \cr
&=& \frac{Am}{r}\left(1+m\frac{d}{dm}\right)\int_0^\infty dp \frac{J_1(pr)}{(p^2+m^2)^{1/2}}  \cr 
&=& \frac{Am}{r}\left(1+m\frac{d}{dm}\right) \left(\frac{1}{mr}-\frac{e^{-mr}}{mr}\right) \cr
&=&Am\frac{e^{-mr}}{r}.
\label{eq:scalar}
\end{eqnarray}
Here, we have used 
\begin{eqnarray}
\int_0^\infty dx \frac{ J_\nu(2ax) }{(x^2+y^2)^{1/2}}
= I_{\nu/2}(ay)K_{\nu/2}(ay)
\label{eq:JIK}
\end{eqnarray}
and its special case of $\nu=1$, $x=p$, $y=m$ and $a=r/2$:
\begin{eqnarray}
&&\int_0^\infty dp \frac{ J_1(pr) }{(p^2+m^2)^{1/2}}
= I_{1/2}(\frac{mr}{2})K_{1/2}(\frac{mr}{2}) \cr
&=& \frac{2}{\sqrt{\pi mr}}\sinh(\frac{mr}{2})\cdot \sqrt{\frac{\pi}{mr}} e^{- mr/2} \cr
&=& \frac{1}{mr}-\frac{e^{-mr}}{mr}.
\end{eqnarray}

Similarly, we obtain 
\begin{eqnarray}
&&\hat s_\mu \hat s_\nu D_{\mu\nu}(s)
= \frac{3}{4\pi^2r^2}
\int_0^\infty dp p\tilde D(p^2) J_2(pr) \cr
&=& \frac{Am}{r^2}
\int_0^\infty dp~ \frac{p}{(p^2+m^2)^{3/2}} J_2(pr) \cr
&=&
-\frac{Am}{r^2}\int_0^\infty dp~\frac{d}{dp} \left[\frac{1}{(p^2+m^2)^{1/2}}\right] J_2(pr) \cr
&=&
\frac{Am}{r^2}\int_0^\infty dp~\frac{1}{(p^2+m^2)^{1/2}} \frac{d}{dp} J_2(pr) \cr
&=&
\frac{Am}{2r}\int_0^\infty dp~\frac{1}{(p^2+m^2)^{1/2}} [J_1(pr)-J_3(pr)] \cr
&=&\frac{Am}{r}\left[\frac{2}{m^3r^3}-\left(\frac{1}{mr}+\frac{2}{m^2r^2}+\frac{2}{m^3r^3}\right)e^{-mr}\right]~~~~~~
\label{eq:tensor}
\end{eqnarray}
using Eq.~(\ref{eq:JIK}) 
with  
$\nu=3$, $x=p$, $y=m$ and $a=r/2$:
\begin{eqnarray}
&&\int_0^\infty dp \frac{ J_3(pr) }{(p^2+m^2)^{1/2}}
= I_{3/2}(\frac{mr}{2})K_{3/2}(\frac{mr}{2}) \cr
&=&\frac{2}{\sqrt{\pi mr}}\left[\cosh(\frac{mr}{2})-\frac{\sinh(\frac{mr}{2})}{mr/2}\right]\cr
&&\times \left(1+\frac{2}{mr}\right)\sqrt{\frac{\pi}{mr}} e^{-mr/2} \cr
&=& \frac{1}{mr}\left(1-\frac{4}{m^2r^2}\right)-\frac{e^{-mr}}{mr}
\left(1+\frac{2}{mr}\right)^2.
\end{eqnarray}
Inserting Eqs.~(\ref{eq:scalar}) and (\ref{eq:tensor}) into Eqs.~(\ref{eq:f-Yukawa}) and (\ref{eq:g-Yukawa}), 
we obtain 
\begin{eqnarray}
f(r)&=& \frac{Am}{3r}\left[e^{-mr}\left(1+\frac{1}{mr}+\frac{2}{m^2r^2}+\frac{2}{m^3r^3}\right)-\frac{2}{m^3r^3}\right] \cr
&=&\frac{Am^2}{3} F(mr),
\label{eq:fr-Yukawa}
\end{eqnarray}
\begin{eqnarray}
g(r)&=&
\frac{Am}{3r}\left[e^{-mr}\left(1+\frac{4}{mr}+\frac{8}{m^2r^2}+\frac{8}{m^3}r^3\right)-\frac{8}{m^3r^3}\right] \cr
&=&\frac{Am^2}{3} G(mr)
\label{eq:gr-Yukawa}
\end{eqnarray}
with $F(w)$ and $G(w)$ 
defined in Eqs.~(\ref{eq:F-fun}) and 
(\ref{eq:G-fun}).

\section{Color-magnetic correlations at each $\beta$}

In Appendix B, we present the color-magnetic correlations for each $\beta$ to provide comprehensive information on lattice QCD data, 
since the scaling property is nontrivial 
for the field-strength correlation.

Figures~\ref{fig:su3perp_each} and \ref{fig:su3para_each} show SU(3)$_{\rm color}$ lattice QCD at each $\beta$ 
for the color-magnetic correlations 
of the perpendicular-type 
$C_{\perp}(r)$ and of the parallel-type 
$C_{\parallel}(r)$, respectively.

Figures~\ref{fig:su2perp_each} and \ref{fig:su2para_each} show SU(2)$_{\rm color}$ lattice QCD at each $\beta$  
for the color-magnetic correlations 
of the perpendicular-type $C_{\perp}(r)$ 
and of the parallel-type 
$C_{\parallel}(r)$, respectively.

\begin{widetext}

\begin{figure}[H]
    \begin{minipage}[b]{0.48\linewidth}
    \centering
    \includegraphics[width=9.2cm]{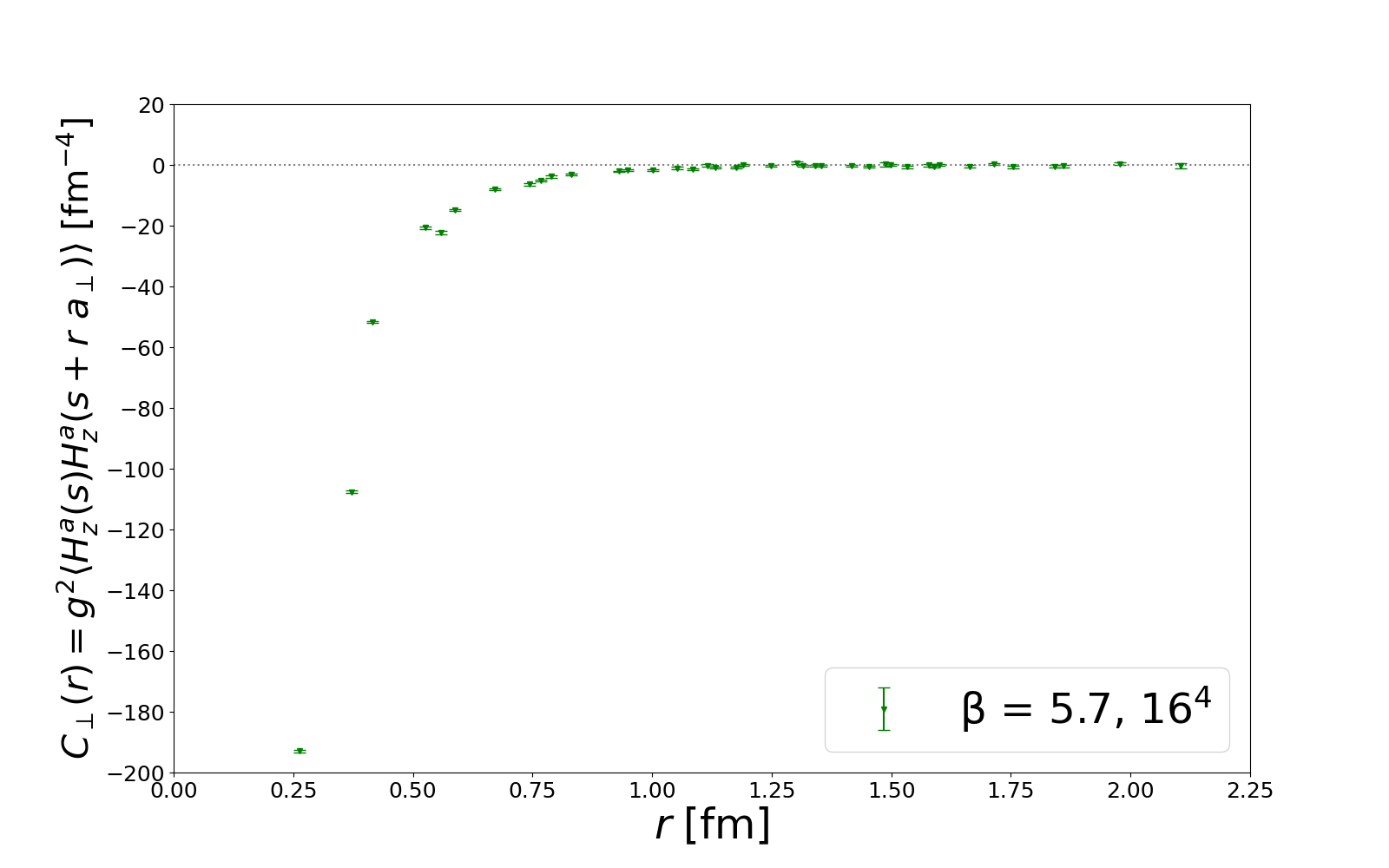}
    \includegraphics[width=9.2cm]{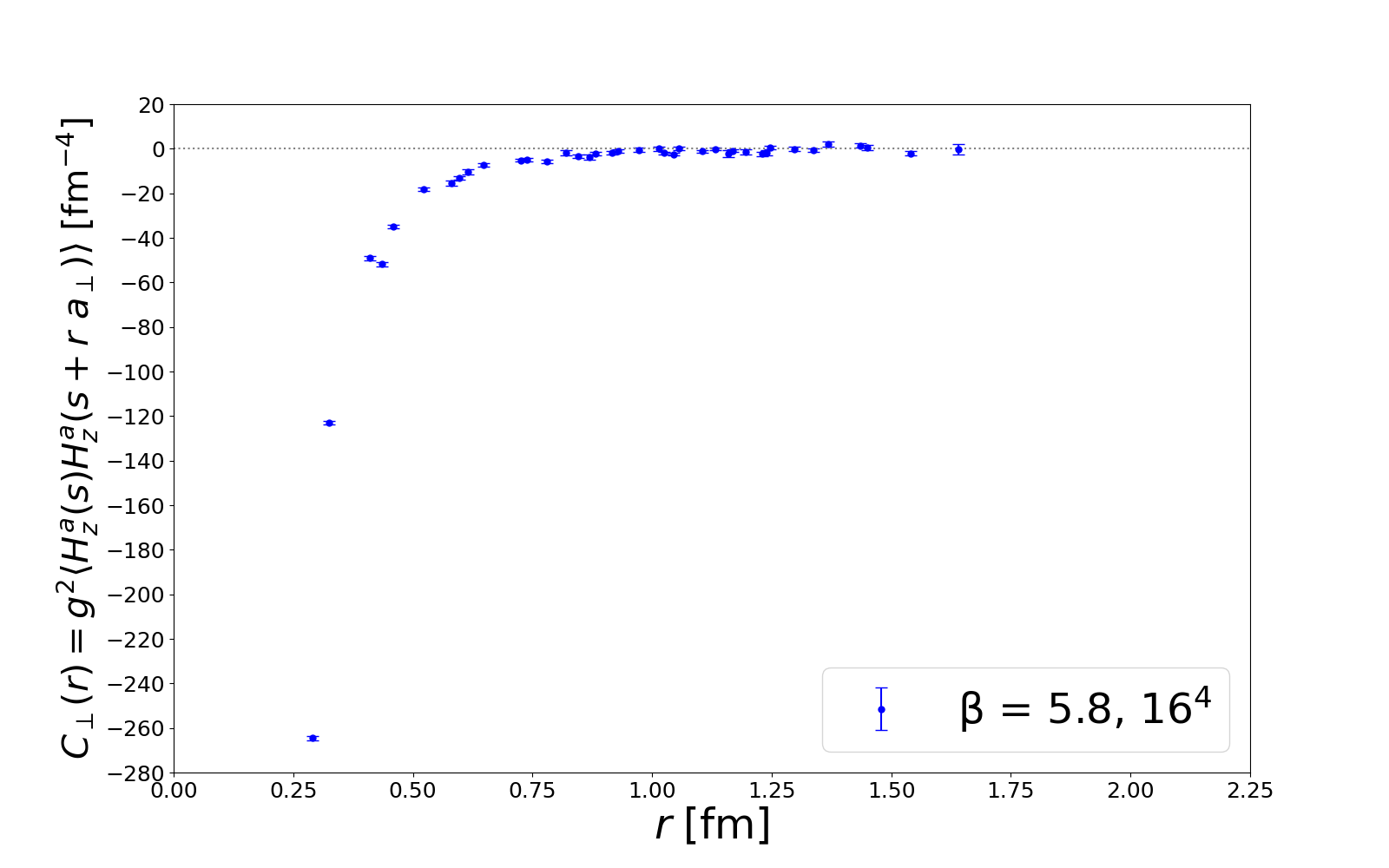}
    \includegraphics[width=9.2cm]{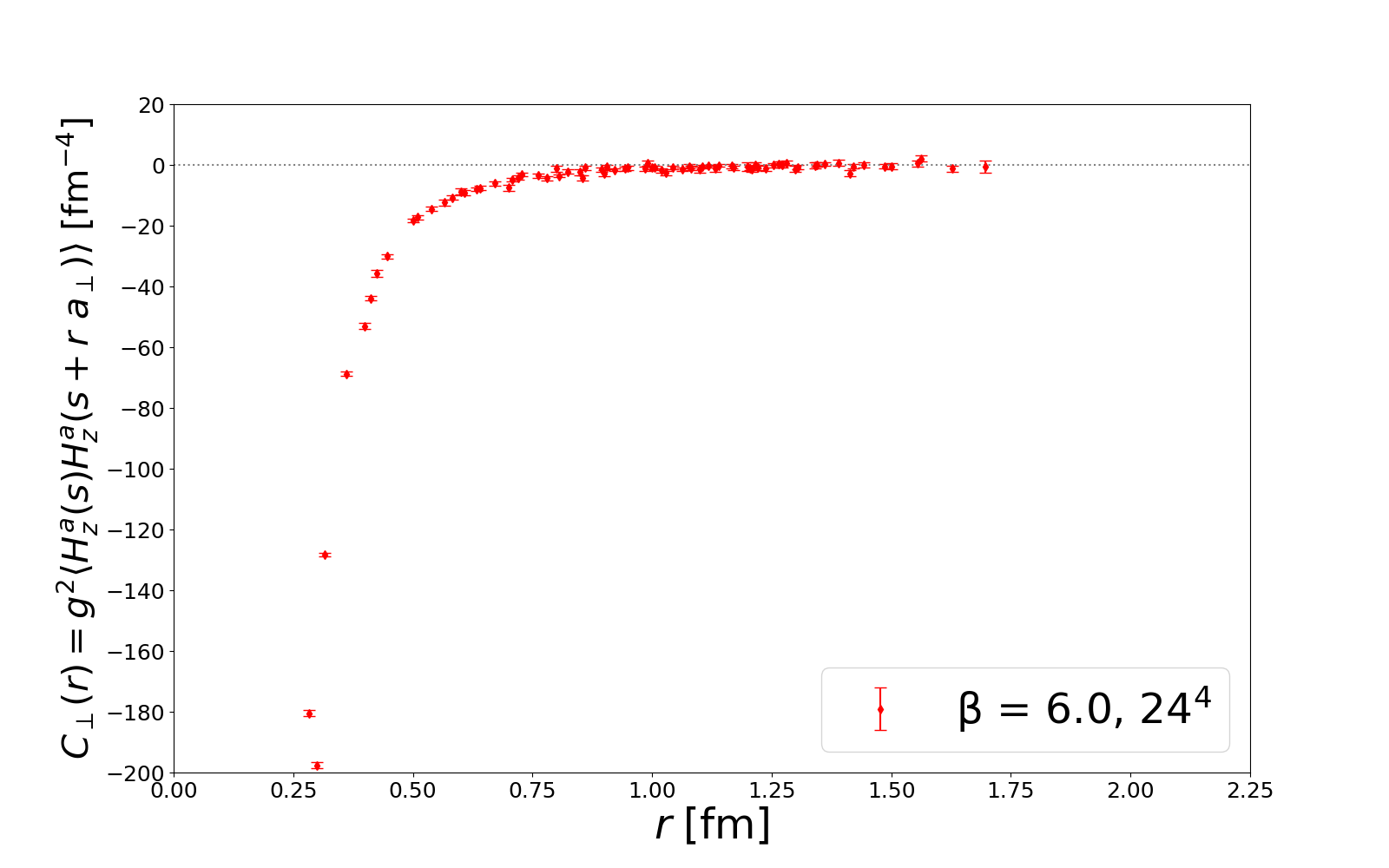}
    \caption{
Perpendicular-type color-magnetic correlations $C_{\perp}(r)$ 
in SU(3)$_{\rm color}$ lattice QCD at 
$\beta$ =5.7, 5.8 and 6.0.
}
    \label{fig:su3perp_each}
\end{minipage}
\begin{minipage}[b]{0.48\linewidth}
    \centering
    \includegraphics[width=9.2cm]{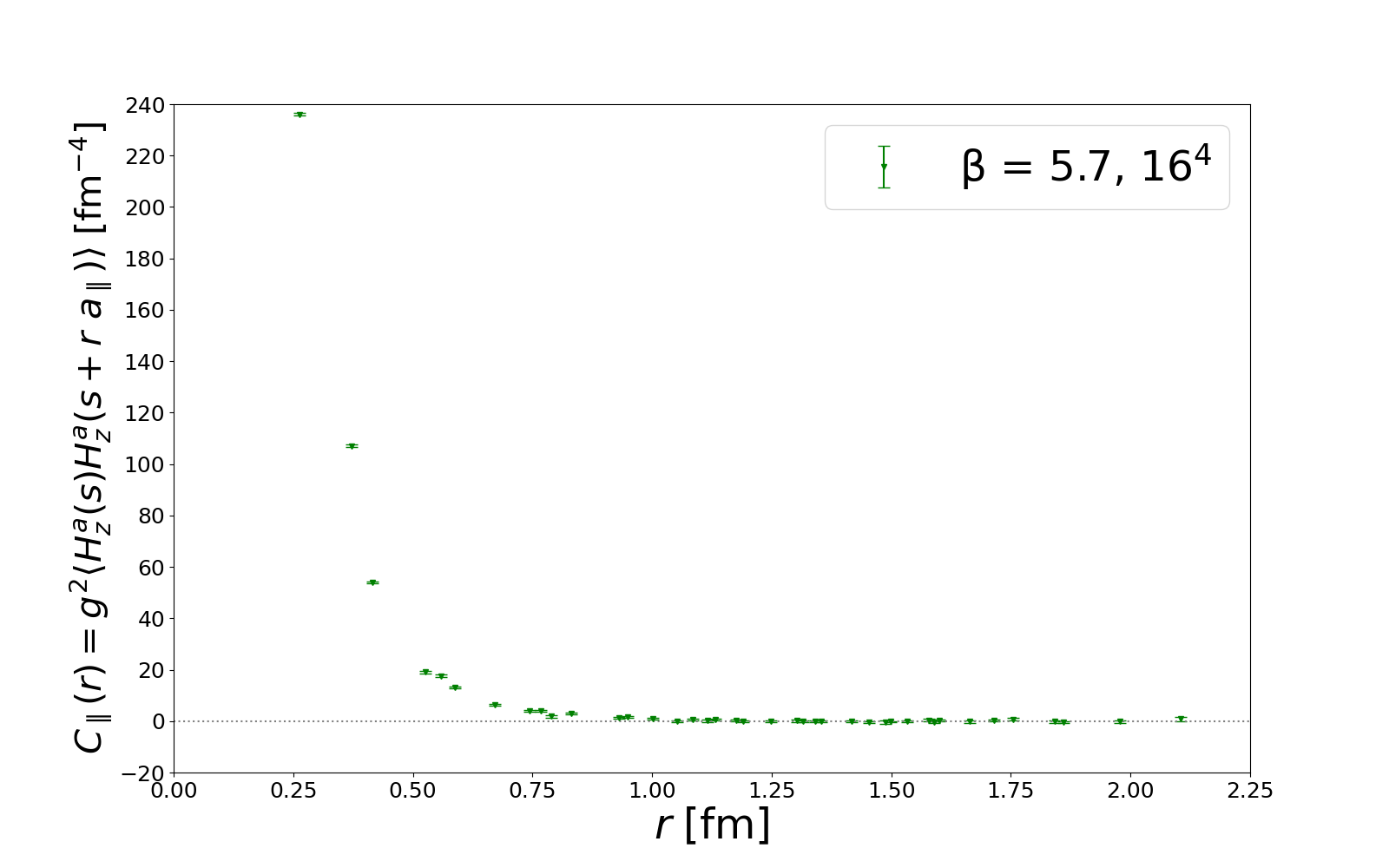}
    \includegraphics[width=9.2cm]{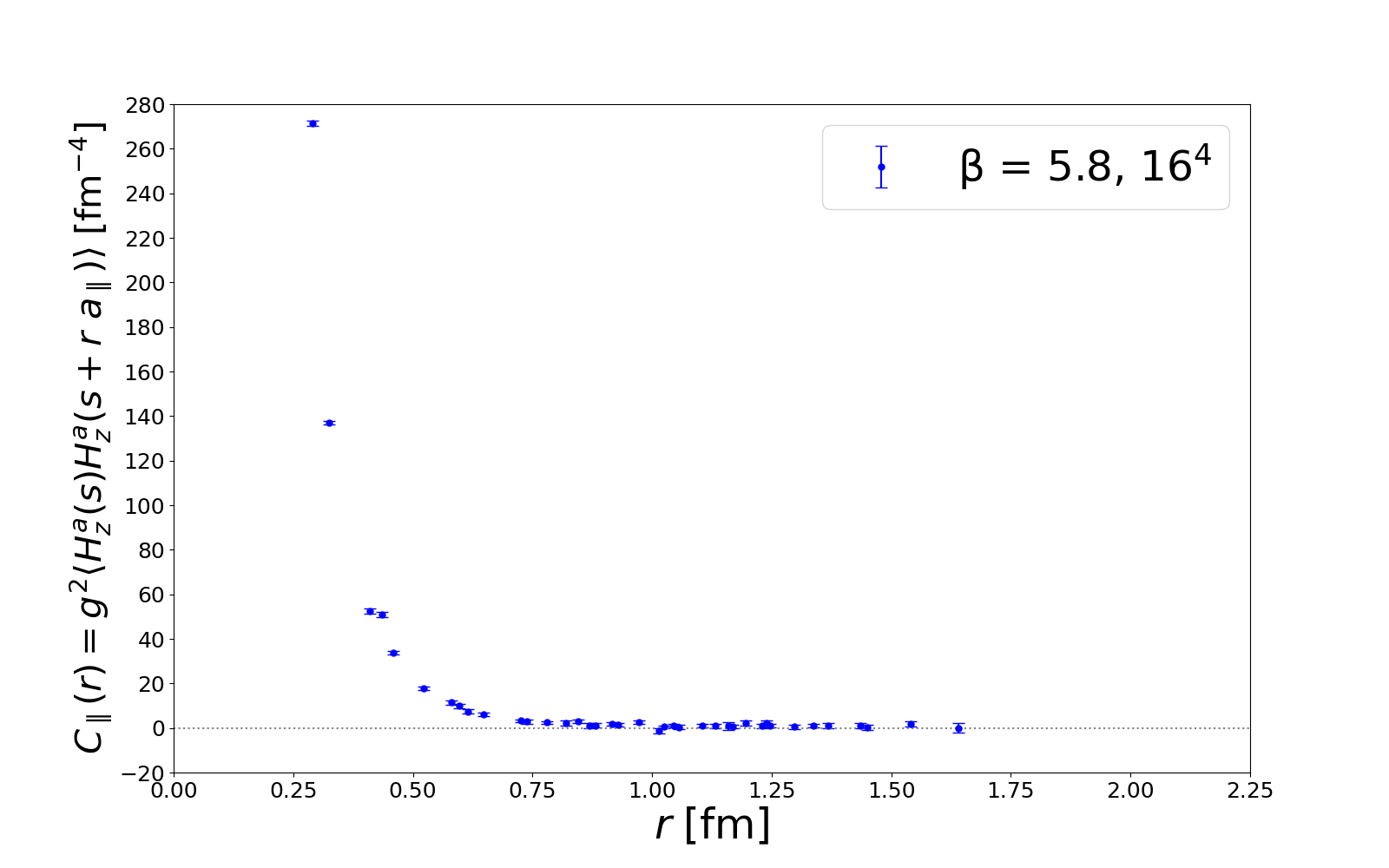}
    \includegraphics[width=9.2cm]{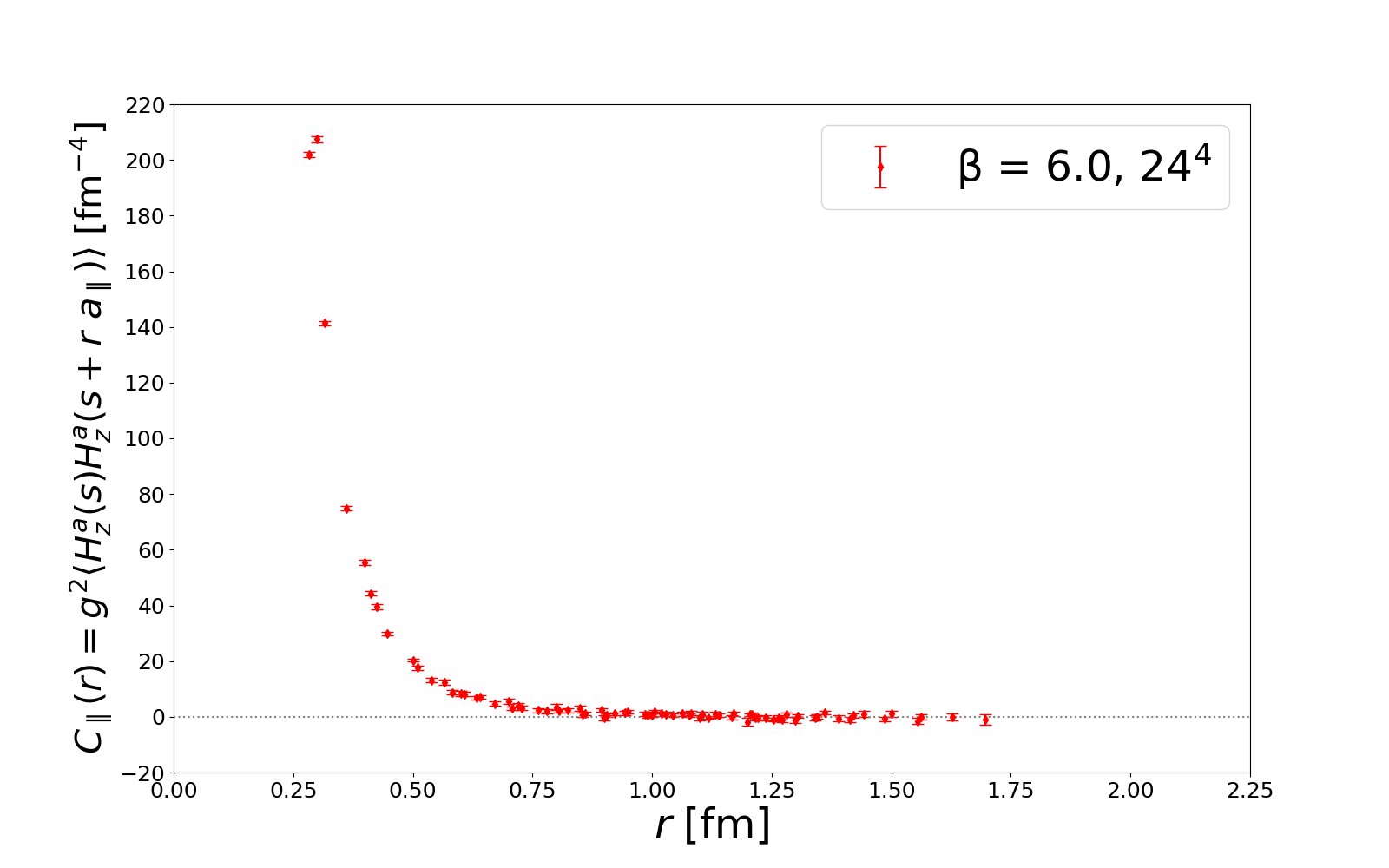}
    \caption{
Parallel-type color-magnetic correlations $C_{\parallel}(r)$ 
in SU(3)$_{\rm color}$ lattice QCD 
at $\beta$=5.7, 5.8 and 6.0.
}
    \label{fig:su3para_each}
    \end{minipage}
\end{figure}

\begin{figure}[H]
    \begin{minipage}[b]{0.48\linewidth}
    \centering
    \includegraphics[width=9.2cm]{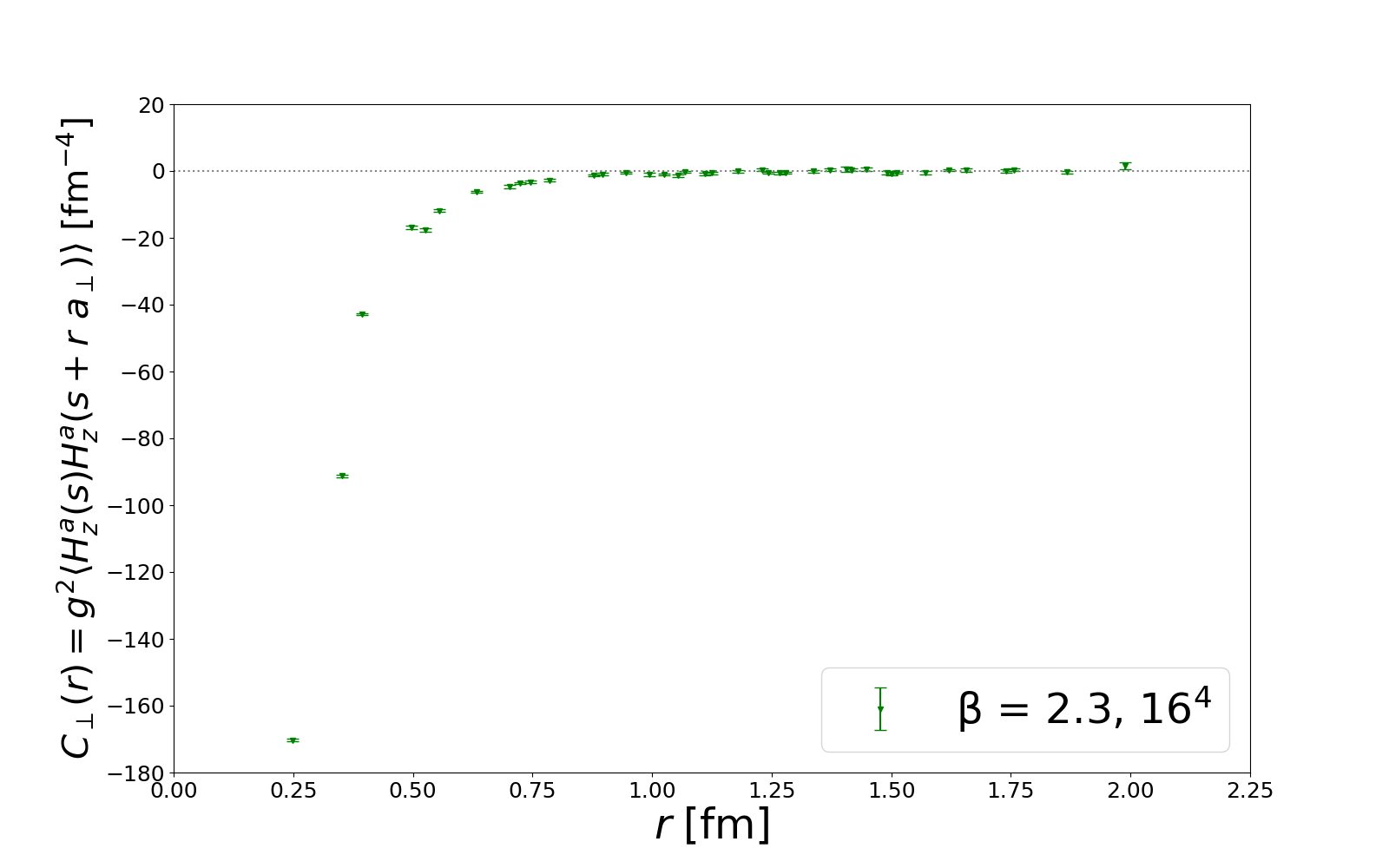}
    \includegraphics[width=9.2cm]{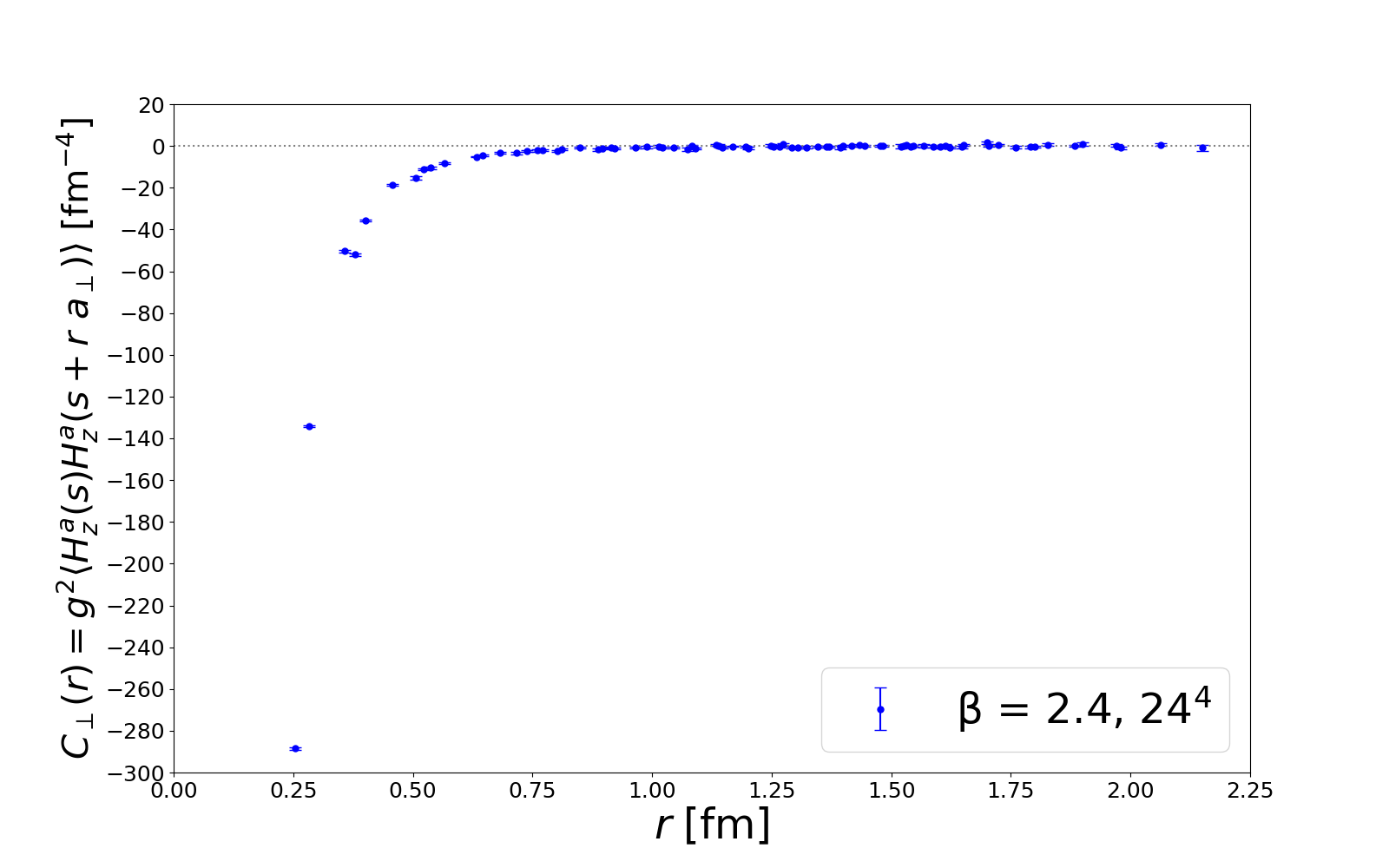}
    \includegraphics[width=9.2cm]{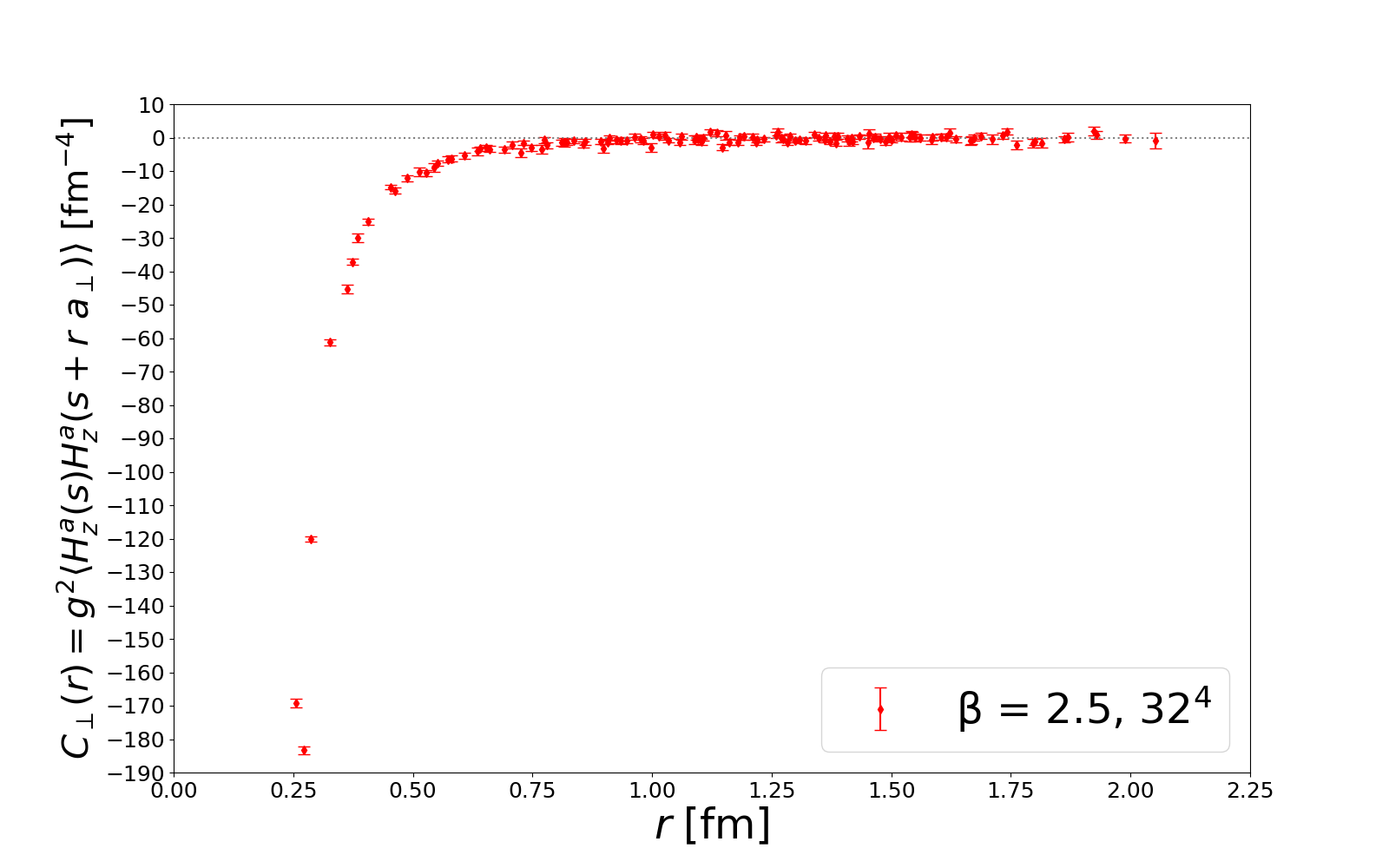}
    \caption{
Perpendicular-type color-magnetic correlations $C_{\perp}(r)$ 
in SU(2)$_{\rm color}$ lattice QCD 
at $\beta$=2.3, 2.4, and 2.5.
}
    \label{fig:su2perp_each}

\end{minipage}
\begin{minipage}[b]{0.48\linewidth}
    \centering
    \includegraphics[width=9.2cm]{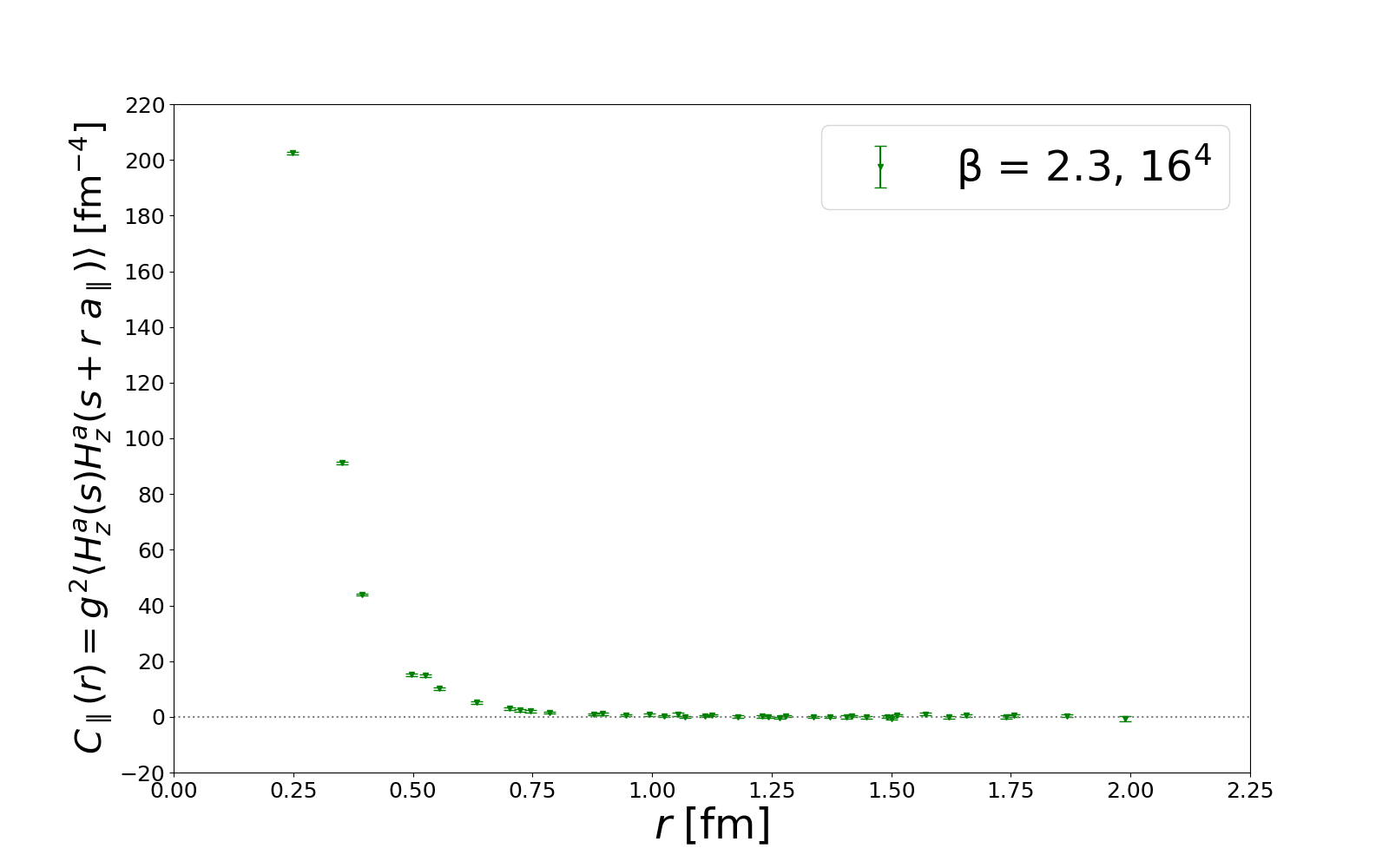}
    \includegraphics[width=9.2cm]{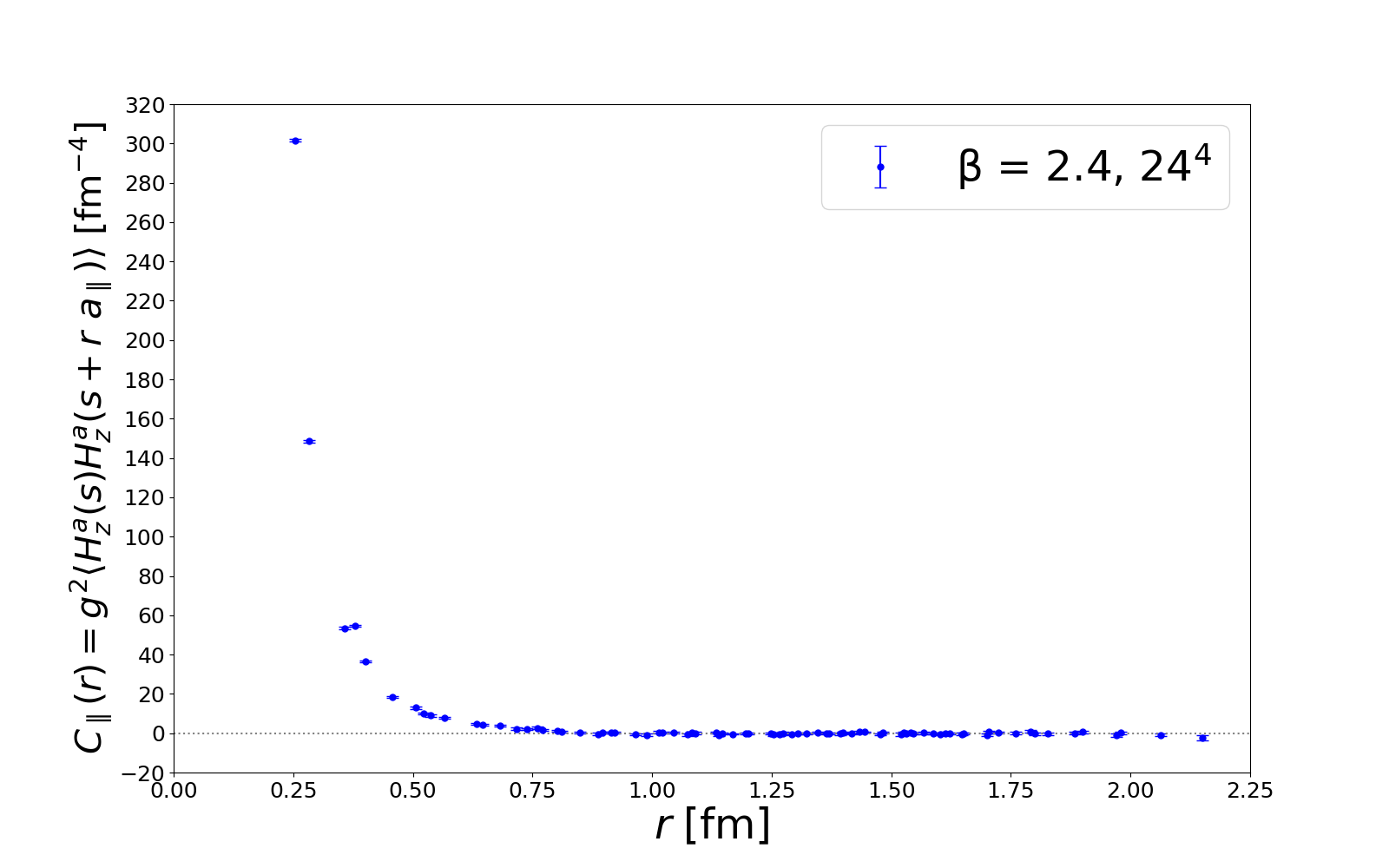}
    \includegraphics[width=9.2cm]{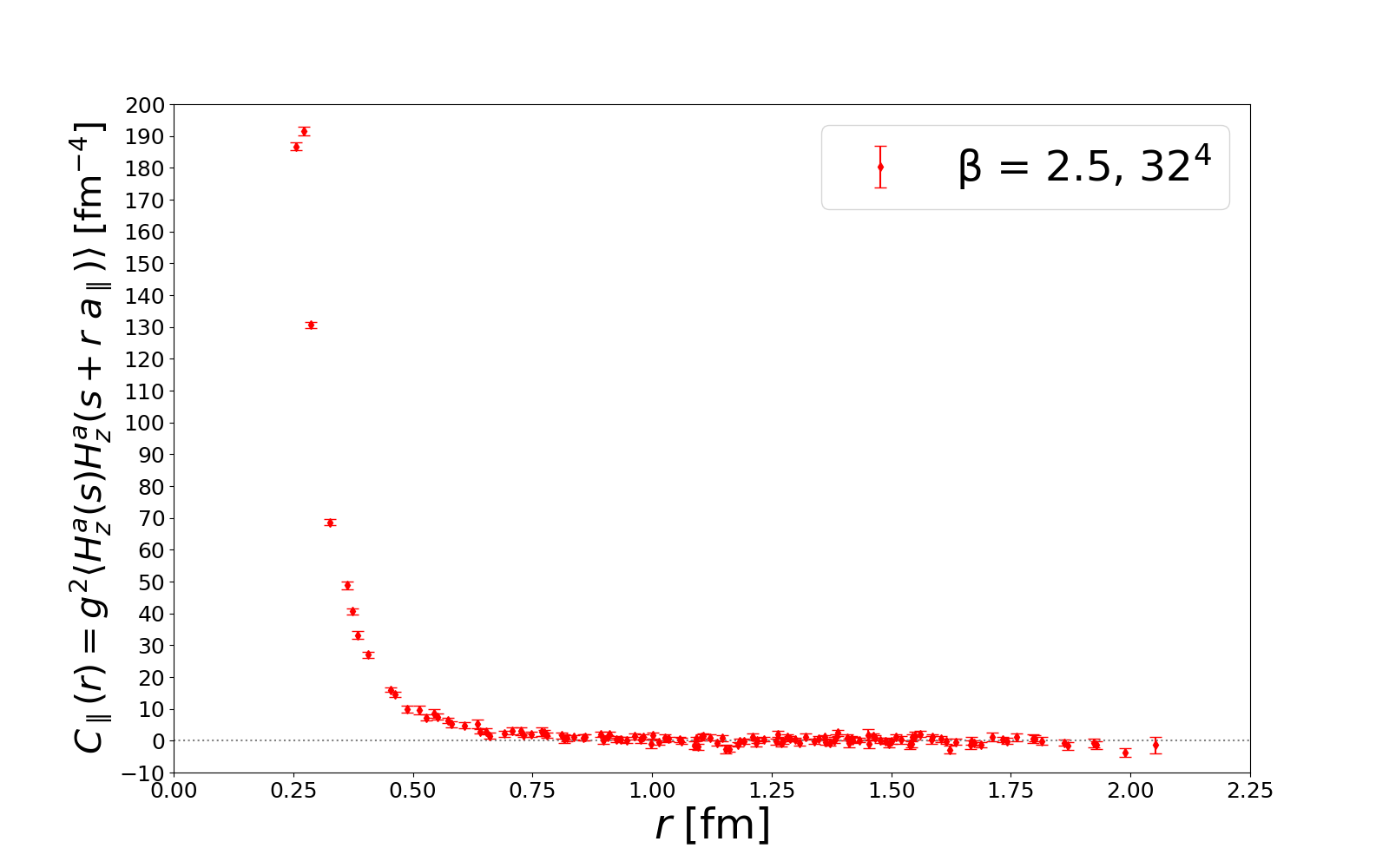}
    \caption{
Parallel-type color-magnetic correlations $C_{\parallel}(r)$ 
in SU(2)$_{\rm color}$ lattice QCD 
at $\beta$=2.3, 2.4, and 2.5.
}
    \label{fig:su2para_each}
    \end{minipage}
\end{figure}
\end{widetext}

\section{Correlation of plaquette field-strength}

\begin{figure}[htbp]
    \centering
\includegraphics[width=9.2cm]{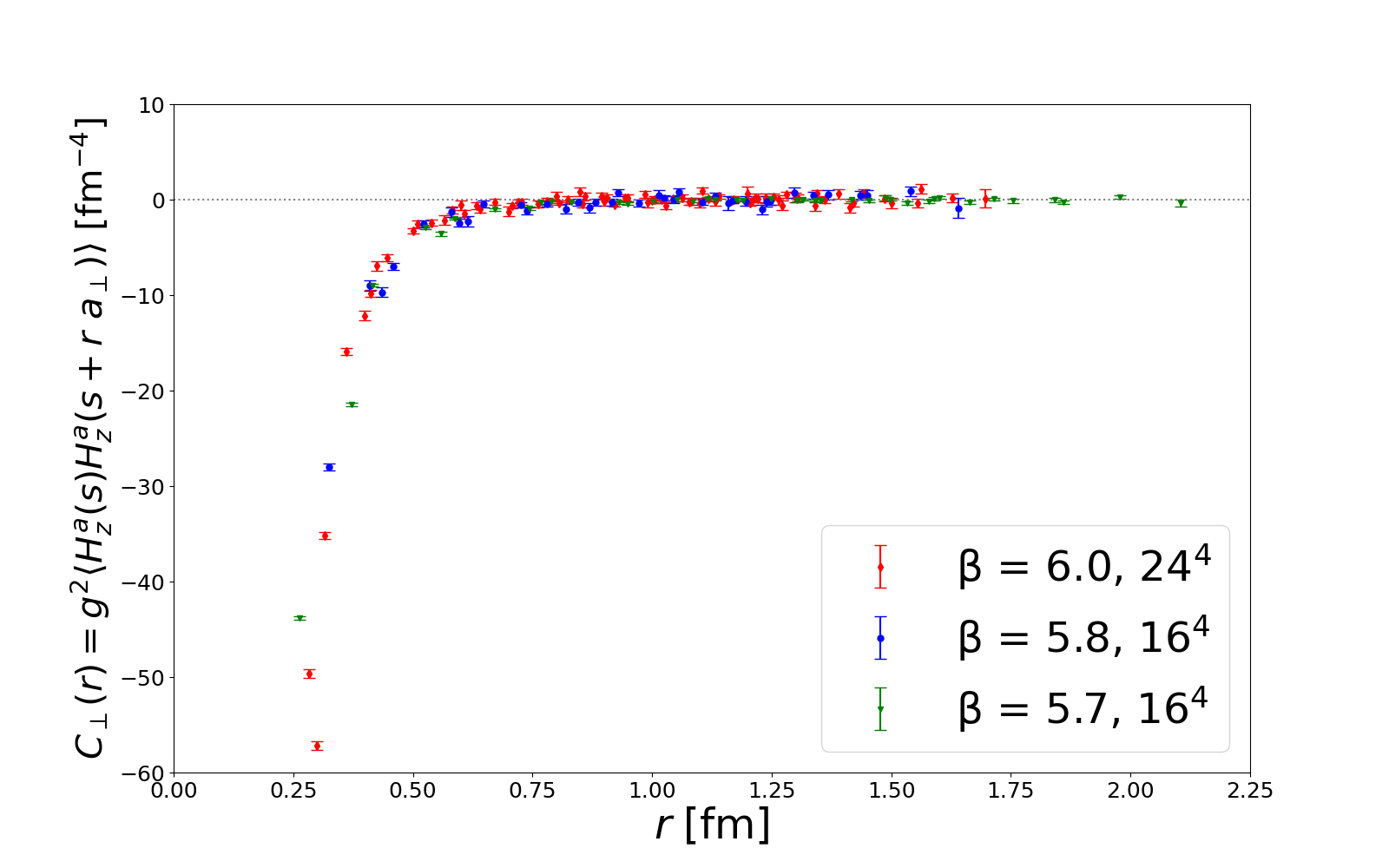}
\includegraphics[width=9.2cm]{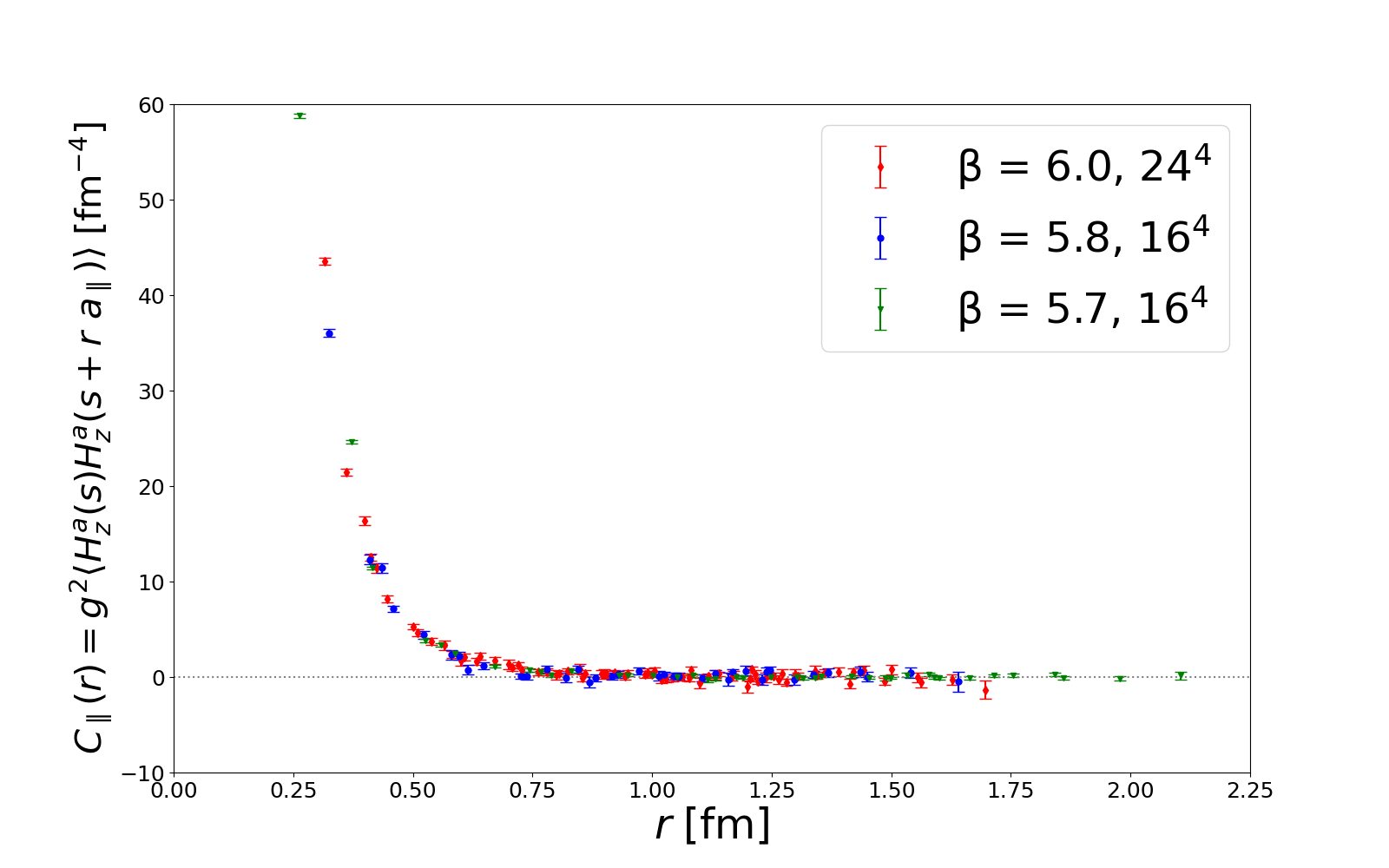}
\includegraphics[width=9.2cm]{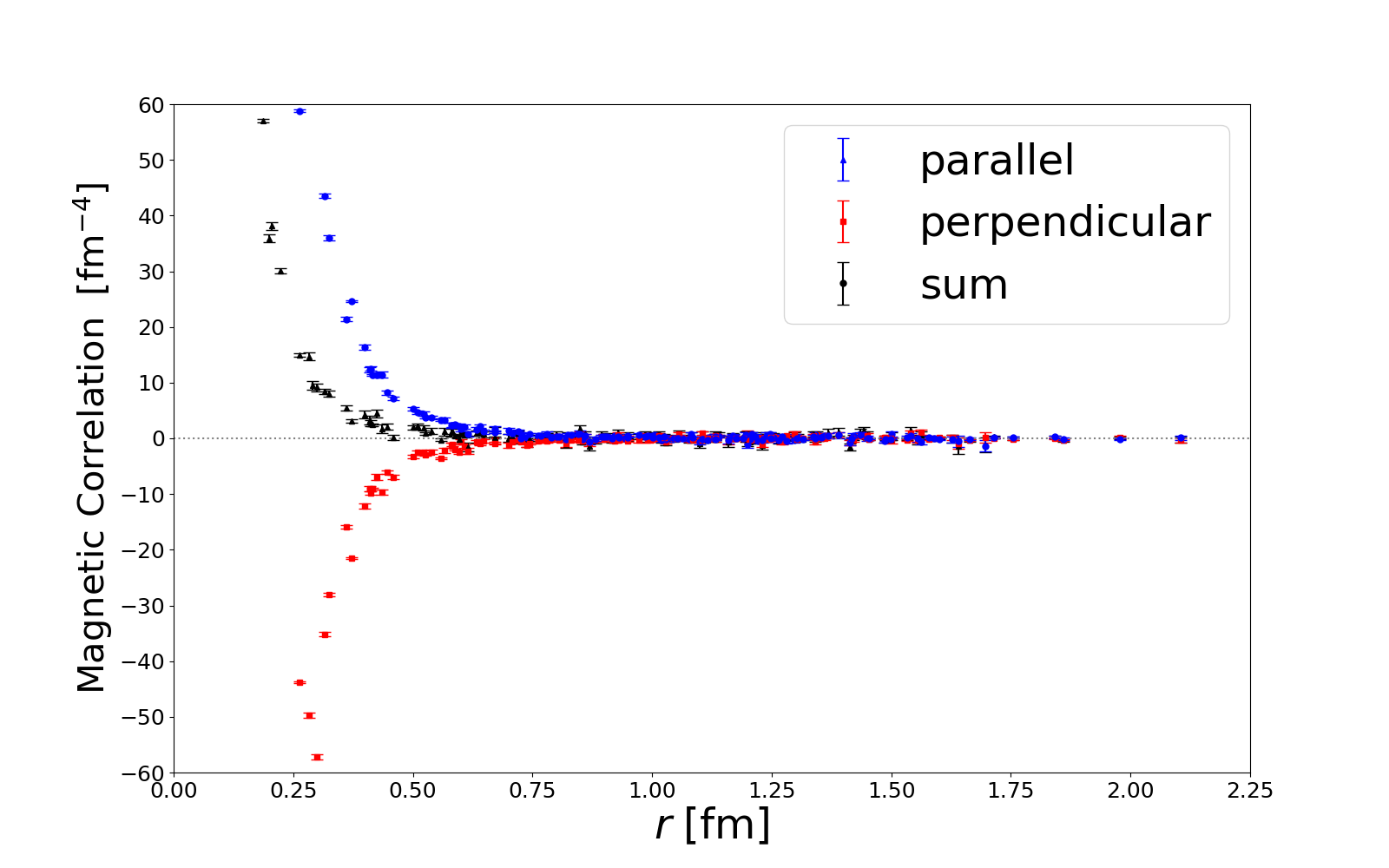}
    \caption{
Color-magnetic correlations using the plaquette field strength ${\cal G}_{\mu\nu}$ in the Landau gauge in SU(3) lattice QCD.
The top shows the perpendicular-type of 
$C_{\perp}^P(r) \equiv g^2\langle {\cal H}^a_z(s){\cal H}^a_z(s+r \hat \perp)\rangle$ ($\hat \perp \in xy\hbox{
-plane}$).
The middle shows the parallel-type of 
$C_{\parallel}^P(r) \equiv g^2\langle {\cal H}^a_z(s){\cal H}^a_z(s+r \hat \parallel)\rangle$ ($\hat \parallel \in zt\hbox{-plane}$).
The bottom shows the sum (black) of 
$C_{\perp}^P(r)$ (red) and  $C_{\parallel}^P(r)$ (blue).
    }
    \label{fig:P-perp}
\end{figure}

In Appendix C, we examine the correlation of the plaquette field strength ${\cal G}_{\mu\nu}(s)$, which is defined in Eq.~(\ref{eq:P-elemag}), 
in the Landau gauge in SU(3) lattice QCD.
We denote the corresponding color-magnetic field by ${\cal H}_i \equiv \epsilon_{ijk} {\cal G}_{jk}(s)/2$ and show the color-magnetic correlation $g^2 \langle {\cal H}_i^a(s) {\cal H}_i^a(s')\rangle$ in Fig.~\ref{fig:P-perp}.

The top of Fig.~\ref{fig:P-perp} shows 
the perpendicular-type color-magnetic correlation, 
$C_{\perp}^P(r) \equiv g^2\langle {\cal H}^a_z(s){\cal H}^a_z(s+r \hat \perp)\rangle$ ($\hat \perp~\in xy\hbox{-plane}$),
which is found to be always negative like $C_{\perp}(r)$, 
except for the same point of $r=0$.
The middle figure shows
the parallel-type correlation,
$C_{\parallel}^P(r) \equiv g^2\langle {\cal H}^a_z(s){\cal H}^a_z(s+r \hat \parallel)\rangle$ ($\hat \parallel~\in zt\hbox{-plane}$),
which is always positive like $C_{\parallel}(r)$.
The bottom figure 
shows the sum of the perpendicular-type and 
parallel-type, $C_{\perp}^P(r)+C_{\parallel}^P(r)$. 
As before, we observe a significant  cancellation between the perpendicular-type and 
parallel-type color-magnetic correlations
in the infrared region of $r \gtrsim 0.4~{\rm fm}$.

\section{Quadratic term in color-magnetic correlation 
in Landau gauge and Yukawa propagator}

In Appendix D, we derive 
the quadratic term in 
the color-magnetic correlation 
in the Landau gauge, 
including the case of the Yukawa-type gluon propagator (\ref{eq:Yukawa-type}).

\subsection{General formalism on quadratic terms in color-magnetic correlation in the Landau gauge}

First, we present the general formalism 
on the quadratic term in the color-magnetic correlation in the Landau gauge  
in four-dimensional Euclidean space-time.

Due to the Lorentz symmetry, 
the Landau-gauge propagator $D_{\mu\nu}(s-s')\equiv g^2\langle A^a_{\mu}(s)A^a_\nu(s')\rangle$ generally takes the form of 
Eq.~(\ref{eq:Landau-gluon}), i.e.,  
$D_{\mu\nu}(s)
=f(r)\delta_{\mu\nu}-g(r)\hat s_\mu \hat s_\nu$,
with $r \equiv |s|$ and $\hat s\equiv s/r$.

The quadratic term in  
the color-magnetic correlation is generally expressed by the propagator 
$D_{\mu\nu}(s)$ as
\begin{eqnarray}
&&g^2\langle H^a_z(s)H^a_z(s')\rangle_{\rm quad}
\equiv g^2\langle \epsilon_{ij3} \partial_i A^a_j(s)
\epsilon_{kl3}\partial_k A^a_l(s') \rangle \cr
&=&g^2
\epsilon_{ij3}\epsilon_{kl3}\partial_i^s \partial_k^{s'}
\langle   A^a_j(s) A^a_l(s') \rangle
\cr
&=&-\epsilon_{ij3}\epsilon_{kl3}\partial_i^s \partial_k^{s}
D_{jl}(s-s').
\end{eqnarray}
Without loss of generality, 
we can set $s=(x,y,z,0)$ and $s'=0$ 
using translation and four-dimensional Euclidean rotation on the $zt$-plane 
to obtain  
\begin{eqnarray}
&&
g^2\langle H^a_z(s)H^a_z(0)\rangle_{\rm quad}
=-\epsilon_{ij3}\epsilon_{kl3}\partial^s_i \partial^s_k
D_{jl}(s) \cr
&=&-\epsilon_{ij3}\epsilon_{kl3}\partial_i \partial_k
[f(r)\delta_{jl}-g(r)
\hat s_j \hat s_l] \cr
&=&-(\delta_{ik}-\delta_{i3}\delta_{k3})\partial_i \partial_k
f(r)+
\epsilon_{ij3}\epsilon_{kl3}\partial_k\left\{g(r)\hat s_j\frac1r \hat \delta_{il}\right\} \cr
&=&-(\partial_x^2+\partial_y^2)f(r)+
\epsilon_{ij3}\epsilon_{ki3}\partial_k\left\{\frac{g(r)}{r}\hat s_j\right\}
\end{eqnarray}
with $r\equiv |s|$,
$\hat s \equiv s/r$, 
$\hat \delta_{ij}\equiv \delta_{ij}-\hat s_i \hat s_j$, and $\partial_i\hat s_j=\frac1r\hat \delta_{ij}$.
Using the cylindrical coordinate $(\rho,\phi,z)$ for $s$, 
we get the quadratic term in the color-magnetic correlation in the Landau gauge: 
\begin{eqnarray}
&&g^2\langle H^a_z(s)H^a_z(0)\rangle_{\rm quad} \cr
&=&-\frac1\rho\frac{\partial}{\partial\rho}\rho \frac{\partial}{\partial \rho} f(r) \cr
&&-
(\delta_{jk}-\delta_{j3}\delta_{k3})\left\{\frac{d}{dr}\left(\frac{g(r)}{r}\right)\hat s_k\hat s_j+\frac{g(r)}{r^2}\hat\delta_{kj}\right\} \cr
&=&-\frac1\rho\frac{\partial}{\partial\rho}\rho \frac{\partial}{\partial \rho} f(r) -
\left\{\frac{d}{dr}\left(\frac{g(r)}{r}\right)+\frac{g(r)}{r^2}\right\} \cr
&&~~~~~~~~~~~~~~~~~~+\hat z^2\left\{\frac{d}{dr}\left(\frac{g(r)}{r}\right) - \frac{g(r)}{r^2}\right\} \cr
&=&-\Big[\frac1\rho\frac{\partial}{\partial\rho}\rho \frac{\partial}{\partial \rho} f(r) +\frac1{r}\frac{d}{dr}g(r) \cr
&&~~~~~~~~~~~~~~~~~~+\frac{z^2}{r^2}\left\{
\frac{2}{r^2}g(r)-\frac{1}{r}\frac{d}{dr}g(r) \right\} \Big].
\label{eq:CMC}
\end{eqnarray}

\subsection{Color-magnetic correlation for Yukawa-type propagator}

For the Yukawa-type propagator (\ref{eq:Yukawa-type}), 
$f(r)$ and $g(r)$ are given by 
Eqs.~(\ref{eq:fr-Yukawa}) and (\ref{eq:gr-Yukawa}), respectively.
Substituting them into Eq.~(\ref{eq:CMC}) yields the following.

In the case of $s=r\hat \perp=(x,y,0,0)$, i.e., $z=0$ and $\rho=r$, corresponding to $C_\perp(r)$, we get
\begin{eqnarray}
&&g^2\langle H^a_z(r\hat \perp)H^a_z(0)\rangle_{\rm quad} \cr
&=&-\Big[\frac1r\frac{d}{dr}r \frac{d}{d r} f(r) +\frac1{r}\frac{d}{dr}g(r) \Big] \cr
&=&-\frac{Am^4}{3}\left.\Big[\frac1w\frac{d}{dw}w \frac{d}{d w} F(w) +\frac1{w}\frac{d}{dw}G(w) \Big]\right|_{w=mr} \cr
&=&-\frac{Am^4}{3}\left.\Big[\frac1w\frac{d}{dw}\Big(w \frac{d}{d w} F(w) +G(w) \Big) \Big]\right|_{w=mr} \cr
&=&+\frac{Am^4}{3} \left.\Big[\frac1w\frac{d}{dw}\Big\{
e^{-w}\Big(1+\frac{1}{w}\Big)
\Big\} \Big]\right|_{w=mr} \cr
&=&-\frac{Am^4}{3} \left. e^{-w} \frac1w 
\Big(1+\frac1w+\frac1{w^2} \Big) 
\right|_{w=mr} <0.
\end{eqnarray}

In the case of $s=r\hat z=(0,0,z,0)$, i.e., $z=r$ and $\rho=0$, corresponding to $C_\parallel(r)$, 
we get
\begin{eqnarray}
&&g^2\langle H^a_z(r\hat z)H^a_z(0)\rangle_{\rm quad} \cr
&=&-\left.\Big[\frac1{\rho}\frac{\partial}{\partial\rho}\rho \frac{\partial r}{\partial \rho} f'(r) +\frac2{r^2}g(r) \Big]\right|_{\rho \rightarrow 0} \cr
&=&-2\Big[\frac1{r}\frac{d}{dr}f(r) +\frac1{r^2}g(r) \Big] \cr
&=&-\frac{2Am^4}{3}\left.\Big[\frac1w\ \frac{d}{d w} F(w) +\frac{1}{w^2}G(w) \Big]\right|_{w=mr} \cr
&=&\frac{2Am^4}{3} \left. e^{-w} \frac1{w^2}
\Big(1+\frac{1}{w}\Big) \right|_{w=mr} >0.
\end{eqnarray}

\end{document}